\documentclass[preprint, preprintnumbers,showpacs, showkeys, aps, prd, amsmath, amsfonts, amssymb,
   eqsecnum, nofootinbib, floatfix, secnumarabic]{revtex4-1}

\usepackage{epsf}
\usepackage{epsfig}
\usepackage{graphics}
\usepackage{graphicx}
\usepackage{amssymb}

\begin{document}
\preprint{CUMQ/HEP 168}

\title{ \Large Neutral Higgs Bosons in the Higgs Triplet Model with nontrivial mixing}

\author {Fatemeh Arbabifar$^{1}$\footnote{Email:
Farbabifar@ipm.ir}}
\author{Sahar Bahrami $^2$\footnote{Email: sahar.bahrami@concordia.ca}}
\author{Mariana Frank$^2$\footnote{Email: mariana.frank@concordia.ca}}

\affiliation{ $^1 $Department of Physics, 
Semnan University, Semnan, Iran, and School of Particles and Accelerators, 
Institute for Research in~Fundamental Sciences (IPM),
P.O. Box 19395-5531, Tehran, Iran,} 
\affiliation{ $^2 $Department of Physics,  
Concordia University, 7141 Sherbrooke St. West ,
Montreal, Quebec, Canada H4B 1R6.}

\date{\today}

\begin{abstract}
We revisit the neutral  Higgs sector of the  Higgs Triplet Model, with non-negligible mixing in the CP-even Higgs sector. We examine the possibility that one of the Higgs boson state is the particle observed at the LHC at 125 GeV, and the other is either the small LEP excess at 98 GeV; or the CMS excess at 136 GeV; or that the neutral Higgs bosons are (almost) degenerate and have both mass 125 GeV. We show that, under general considerations, an (unmixed) neutral Higgs boson cannot have an enhanced decay branching ratio into $\gamma \gamma$ with respect to the Standard Model one.  An enhancement is however possible for the mixed case, but only for the heavier of the two neutral Higgs bosons, and not for  mass-degenerate Higgs bosons. At the same time the branching ratios into $WW^*,~ZZ^*, ~b {\bar b}$ and $\tau ^+\tau^-$ are similar to the Standard Model, or reduced. We correlate the branching ratios of both Higgs states  into  $Z \gamma$  to those into $\gamma \gamma$ for the three scenarios. The mixed neutral sector of the Higgs triplet model exhibits some  features which could distinguish it from other scenarios at the LHC.

\end{abstract}
\maketitle

 \section{Introduction}
 \label{sec:intro}

 The hunt for Higgs in the Standard Model (SM) and beyond has been given a big boost with the recently discovered resonance at $\simeq 125-126$ GeV,  observed by ATLAS \cite{:2012gk} and CMS \cite{:2012gu} at $5 \sigma$. While this particle resembles in most features the SM Higgs boson, 
 the data hints of enhancements in the $\gamma \gamma$ event rates (although this signal is, at about 2$\sigma$, not sufficiently statistically secure), as well as 
 depressed rates into $\tau^+ \tau^-$ and  $WW^*$, which could hint at extended symmetries. The signals are also consistent with the  findings at the Tevatron \cite{Aaltonen:2012qt}.
 If the $\gamma \gamma$ signals persist with more statistics, they would be an encouraging sign of physics beyond the Standard Model (BSM). This possibility has already inspired many explorations in literature \cite{enhancedphoton}. The decay into $\gamma \gamma$ is loop-induced and thus, sensitive to new physics contributions. The simplest explanation would be the presence of a charged boson in the loop, most likely a charged Higgs boson which appears in most BSM scenarios. In addition, if taken at face value, the suppression of the leptonic modes could be an indication that the neutral boson observed is not a pure SM state, but a mixed state, in which the other component has depressed couplings to leptons.
 
 There are additional hints that  more than one Higgs bosons might have been observed. For instance, CMS observes an additional excess in $\gamma \gamma$ and $\tau$ channel at $\sim 136$ GeV \cite{CMS136}, which also seems to provide a best fit to the Tevatron data \cite{:2012zzl}. Additionally, LEP has observed an excess in $e^+e^- \to Z b {\bar b}$ near $\sim 98$ GeV \cite{Schael:2006cr,Barate:2003sz}. This has lead several authors \cite{Carena:2012xa,Chang:2012ta,Chiang:2012qz,Dorsner:2012pp,Belanger:2012sd,Belanger:2012tt} to investigate the possibility that the data could be fit by not one, but two Higgs bosons; or two degenerate, or nearly degenerate,   Higgs bosons \cite{Ferreira:2012nv}.  
 
 Motivated by these observations, we investigate one of the simplest extensions of the SM, the Higgs Triplet Model (HTM) with nontrivial mixing in the neutral sector. We probe whether the CP-even Higgs states can explain the signal at 125 GeV, and either the additional state at 98 GeV, or the one at 136 GeV.  The HTM has two important ingredients lacking in the SM.  First, it provides an explanation for small neutrino masses \cite{Cheng:1980qt,Schechter:1980gr} through the seesaw mechanism \cite{Mohapatra:1980yp}: even if the boson at the LHC turns out to be completely consistent with the SM boson,   the SM leaves the question of neutrino masses unresolved.  Second, the model includes in its Higgs spectrum one singly-charged and doubly charged boson, making loop-enhancements of decays into $\gamma \gamma$ possible.
 
 The neutral Higgs sector of the HTM has been studied previously \cite{Akeroyd:2007zv,Akeroyd:2009nu,Fukuyama:2009xk,Petcov:2009zr,Fukuyama:2010mz,Akeroyd:2010ip,Arhrib:2011uy,Aoki:2011pz,Kanemura:2012rs,Arhrib:2011vc,Akeroyd:2005gt}. Various authors have provided analyses showing the $\gamma \gamma$ signal suppressed with respect to the SM \cite{Aoki:2011pz, Arhrib:2011vc}. An exception to this is in \cite{Akeroyd:2012ms}, where it was shown that in the case where the triple boson coupling is negative, the decay rates to $\gamma \gamma$ are enhanced.
 
 But most of the authors have considered the HTM for the case in which the mixing in the CP-even neutral bosons is negligible, with the exception of \cite{Akeroyd:2010je}, where the mixing is assumed to be maximal. For negligible mixing, the neutral boson visible at LHC is SM-like (a neutral component of a doublet Higgs representation) with the same tree-level couplings to fermions and gauge bosons, but which also couples  to singly and doubly charged Higgs bosons, possibly a source on enhancement for the $\gamma \gamma$ signal. The production cross sections and decays to $f {\bar f},~WW^*$ and $ZZ^*$ are unchanged with respect to the SM. Should these rates be different, new particles must be added to the model to provide a viable explanation \cite{Wang:2012gm}.
 
 We revisit the model for the case where the mixing is non-negligible, and both states are mixtures of doublet and triplet Higgs representations. We study the tree-level and loop induced ($\gamma \gamma$ and $Z\gamma$) decays of the two bosons  for the case in which $m_{H_1}=125$ GeV, $m_{H_2}=136$ GeV (motivated by the CMS data);  for the case where $m_{H_1}=98$ GeV, $m_{H_2}=125$ GeV (motivated by the LEP excess); and for the case where the two Higgs are degenerate in mass and $m_{H_1}=m_{H_2}=125$ GeV, which is motivated by the case where there is a single boson observed at the LHC. We do not assume specific mixing, but rather study the variation in all parameters due to mixing, and  comment on the case where the mixing is negligible as a limiting case. We consider deviations from unity of ratios of branching ratios 
  in the HTM (with $H_1,\,H_2$ Higgs bosons) versus the SM (with $\Phi$ Higgs boson): 
\begin{equation}
R_{H_1,H_2 \to XX}=\frac{[\sigma( gg \to  H_1,H_2)\times BR(H_1,H_2 \to XX)]_{HTM}}{[\sigma( gg \to \Phi) \times BR(\Phi \to XX )]_{SM}}
\end{equation}
with $XX=\gamma \gamma, f {\bar f}, ZZ^\star, WW^\star$, and predict the rate for $Z\gamma$, as the correlation between this decay and $\gamma\gamma$ would be a further test of the structure of the model.

Our paper is organized as follows: in Section \ref{sec:model} we summarize the main features of the HTM, paying particular attention to the neutral Higgs sector, and outline the conditions on the relevant parameters.  In Section \ref{sec:decays} we present expressions for the decay width of the neutral Higgs bosons, as well as give general analytic expressions for the decay rates, for both $\gamma \gamma$ in \ref{sec:gamgam} and, following examination of the effect of the total width difference between the Higgs boson in SM and in the HTM in \ref{sec:width}, the tree-level decays to $f\bar f$, $WW^*$ and $ZZ^*$ in \ref{sec:tree}.  We follow in Section \ref{sec:analysis} with numerical analysis for the decays of the bosons in scenarios inspired by the experimental data. In Section \ref{sec:Zgamma} we show predictions for the same parameter space for the decays of the neutral Higgs bosons to $Z \gamma$, another indicator of extra charged particles in the model. We summarize our findings and conclude in Section \ref{sec:conclusion}. 

\section{The Higgs Triplet Model}
\label{sec:model}

We briefly describe, for completeness,  the Higgs Triplet Model (HTM), which has been recently the topic  extensive studies  \cite{Arhrib:2011vc,Arhrib:2011uy,Aoki:2011pz,Kanemura:2012rs}.
The  HTM is based on the same symmetry group as the SM, $SU(2)_L \times U(1)_Y$. The only difference is the addition of one triplet field $\Delta$ with hypercharge $Y=1$ to the SM Higgs sector, which already contains one isospin doublet field $\Phi$ with 
hypercharge $Y=1/2$ and  the lepton number $L=2$.  
The relevant terms in Lagrangian are: 
\begin{eqnarray}
\mathcal{L}_{\rm{HTM}}=\mathcal{L}_{\rm{kin}}+\mathcal{L}_{Y}-V(\Phi,\Delta), 
\end{eqnarray}
where $\mathcal{L}_{\rm{kin}}$, $\mathcal{L}_{Y}$ and $V(\Phi,\Delta)$ are 
the kinetic term, Yukawa interaction and the scalar potential, respectively. 
The kinetic term of the Higgs fields is 
\begin{eqnarray}
\mathcal{L}_{\rm{kin}}&=&(D_\mu \Phi)^\dagger (D^\mu \Phi)+\rm{Tr}[(D_\mu \Delta)^\dagger (D^\mu \Delta)], 
\end{eqnarray}
with  
\begin{equation}
D_\mu \Phi=\left(\partial_\mu+i\frac{g}{2}\tau^aW_\mu^a+i\frac{g'}{2}B_\mu\right)\Phi, \quad
D_\mu \Delta=\partial_\mu \Delta+i\frac{g}{2}[\tau^aW_\mu^a,\Delta]+ig'B_\mu\Delta, 
\end{equation}
the covariant derivatives for the doublet and triplet Higgs fields.
The Yukawa interaction for the Higgs fields is  
\begin{eqnarray}
\mathcal{L}_Y&=&-\left[\bar{Q}_L^iY_d^{ij}\Phi d_R^j+\bar{Q}_L^iY_u^{ij}\tilde{\Phi}u_R^j+\bar{L}_L^iY_e^{ij}\Phi e_R^j+\rm{h.c.}\right] + h_{ij}\overline{L_L^{ic}}i\tau_2\Delta L_L^j+\rm{h.c.},~~~~ \label{nu_yukawa}
\end{eqnarray}
where $\tilde{\Phi}=i\tau_2 \Phi^*$,  $Y_{u,d,e}$ are  3$\times$3 complex matrices, and $h_{ij}$ is a $3\times 3$ complex symmetric Yukawa matrix. 
The most general Higgs potential involving the doublet $\Phi$ and triplet $\Delta$  is given by 
\begin{eqnarray}
V(\Phi,\Delta)&=&m^2\Phi^\dagger\Phi+M^2\rm{Tr}(\Delta^\dagger\Delta)+\left[\mu \Phi^Ti\tau_2\Delta^\dagger \Phi+\rm{h.c.}\right]+\lambda_1(\Phi^\dagger\Phi)^2 \nonumber\\
&+&\lambda_2\left[\rm{Tr}(\Delta^\dagger\Delta)\right]^2 +\lambda_3\rm{Tr}[(\Delta^\dagger\Delta)^2]
+\lambda_4(\Phi^\dagger\Phi)\rm{Tr}(\Delta^\dagger\Delta)+\lambda_5\Phi^\dagger\Delta\Delta^\dagger\Phi,~~~~ \label{pot_htm}
\end{eqnarray}
where $m$ and $M$ are the Higgs bare masses, $\mu$ is the lepton-number violating  parameter,    and 
$\lambda_1$-$\lambda_5$ are Higgs coupling constants. We assume all the parameters to be real.
The scalar fields $\Phi$ and $\Delta$ can written as:
\begin{eqnarray}
\Phi=\left[
\begin{array}{c}
\varphi^+\\
\frac{1}{\sqrt{2}}(\varphi+v_\Phi+i\chi)
\end{array}\right],\quad \Delta =
\left[
\begin{array}{cc}
\frac{\Delta^+}{\sqrt{2}} & \Delta^{++}\\
\frac{1}{\sqrt{2}}(\delta+v_\Delta+i\eta) & -\frac{\Delta^+}{\sqrt{2}} 
\end{array}\right],
\end{eqnarray}
where $v_\Phi$ and $v_\Delta$ 
are the VEVs of the doublet Higgs field and the triplet Higgs field, with 
$v^2\equiv v_\Phi^2+2v_\Delta^2\simeq$ (246 GeV)$^2$. The electric charge is defined as $Q=I_{3}+Y$, with $I_{3}$ the third component of the $SU(2)_L$ isospin.

Minimizing the potential  with respect to the VEVs $v_\Phi,\, v_\Delta$ yields expressions for $m, M$ in terms of the other coefficients in the model.  
 The mass matrices for the Higgs bosons are diagonalized by unitary matrices, yielding physical states for the singly charged, the CP-odd, and the CP-even neutral scalar sectors, respectively:
\begin{eqnarray}
\left(
\begin{array}{c}
\varphi^\pm\\
\Delta^\pm
\end{array}\right)&=&
\left(
\begin{array}{cc}
\cos \beta_\pm & -\sin\beta_\pm \\
\sin\beta_\pm   & \cos\beta_\pm
\end{array}
\right)
\left(
\begin{array}{c}
w^\pm\\
H^\pm
\end{array}\right),\quad 
\left(
\begin{array}{c}
\chi\\
\eta
\end{array}\right)=
\left(
\begin{array}{cc}
\cos \beta_0 & -\sin\beta_0 \\
\sin\beta_0   & \cos\beta_0
\end{array}
\right)
\left(
\begin{array}{c}
z\\
A
\end{array}\right),\nonumber\\
\left(
\begin{array}{c}
\varphi\\
\delta
\end{array}\right)&=&
\left(
\begin{array}{cc}
\cos \alpha & -\sin\alpha \\
\sin\alpha   & \cos\alpha
\end{array}
\right)
\left(
\begin{array}{c}
h\\
H
\end{array}\right),
\end{eqnarray}
where the mixing angles are  in the same sectors are given by  
\begin{eqnarray}
\tan\beta_\pm&=&\frac{\sqrt{2}v_\Delta}{v_\Phi},\quad \tan\beta_0 = \frac{2v_\Delta}{v_\Phi}, \nonumber\\
\tan2\alpha &=&\frac{v_\Delta}{v_\Phi}\frac{2v_\Phi^2(\lambda_4+\lambda_5)-4M_\Delta^2}{2v_\Phi^2\lambda_1-M_\Delta^2-2 v_\Delta^2(\lambda_2+\lambda_3)}.~~~~~~~~ \label{tan2a}
\end{eqnarray}
where  $\displaystyle M_\Delta^2\equiv \frac{v_\Phi^2\mu}{\sqrt{2}v_\Delta}$.  
There are seven physical mass eigenstates $H^{\pm\pm}$, $H^\pm$, $A$, $H$ and $h$, in addition to the three Goldstone bosons $w^\pm$ and $z$ which give mass to the gauge bosons. 
The masses of the physical states are expressed in terms of the parameters in the Lagrangian as 
\begin{eqnarray}
&&m_{H^{++}}^2=M_\Delta^2-v_\Delta^2\lambda_3-\frac{\lambda_5}{2}v_\Phi^2,\label{mhpp}\\
&&m_{H^+}^2=\left(M_\Delta^2-\frac{\lambda_5}{4}v_\Phi^2\right)\left(1+\frac{2v_\Delta^2}{v_\Phi^2}\right),\label{mhp}\\
&&m_A^2 =M_\Delta^2\left(1+\frac{4v_\Delta^2}{v_\Phi^2}\right), \label{mA}\\
&&m_h^2= 2v_\Phi^2\lambda_1\cos^2\alpha+\left [M_\Delta^2+2v_\Delta^2(\lambda_2+\lambda_3)\right ] \sin^2\alpha+\left [\frac{2v_\Delta}{v_\Phi}M_\Delta^2-v_\Phi v_\Delta(\lambda_4+\lambda_5)\right] \sin2\alpha,\nonumber\\
&&\\
&&m_H^2=2v_\Phi^2\lambda_1\sin^2\alpha+ \left [M_\Delta^2+2v_\Delta^2(\lambda_2+\lambda_3)\right] \cos^2\alpha- \left [\frac{2v_\Delta}{v_\Phi}M_\Delta^2-v_\Phi v_\Delta(\lambda_4+\lambda_5)\right ] \sin2\alpha,\nonumber\\
\end{eqnarray}
Conversely, the  six parameters $\mu$ and $\lambda_1$-$\lambda_5$ in the Higgs potential 
can be written in terms of the physical scalar masses, the mixing angle $\alpha$ and doublet and triplet VEVs $v_\Phi$ and $v_\Delta$: 
\begin{eqnarray}
\label{eq:lambdavalues}
\mu&=&\frac{\sqrt{2}v_\Delta^2}{v_\Phi^2}M_\Delta^2 =\frac{\sqrt{2}v_\Delta}{v_\Phi^2+4v_\Delta^2}m_A^2,\\
\lambda_1 & = &\frac{1}{2v_\Phi^2}(m_h^2\cos^2\alpha+m_H^2\sin^2\alpha),\\
\lambda_2 & =& \frac{1}{2v_\Delta^2}\left[2m_{H^{++}}^2+v_\Phi^2\left(\frac{m_A^2}{v_\Phi^2+4v_\Delta^2}-\frac{4m_{H^+}^2}{v_\Phi^2+2v_\Delta^2}\right)
+m_H^2\cos^2\alpha+m_h^2\sin^2\alpha\right], \nonumber\\
 \\
\lambda_3 & =& \frac{v_\Phi^2}{v_\Delta^2}\left(\frac{2m_{H^+}^2}{v_\Phi^2+2v_\Delta^2}-\frac{m_{H^{++}}^2}{v_\Phi^2}-\frac{m_A^2}{v_\Phi^2+4v_\Delta^2}\right),  \\
\lambda_4 & =& \frac{4m_{H^+}^2}{v_\Phi^2+2v_\Delta^2}-\frac{2m_A^2}{v_\Phi^2+4v_\Delta^2}+\frac{m_h^2-m_H^2}{2v_\Phi v_\Delta}\sin2\alpha,  \\
\lambda_5 & =& 4\left(\frac{m_A^2}{v_\Phi^2+4v_\Delta^2}-\frac{m_{H^+}^2}{v_\Phi^2+2v_\Delta^2}\right).
\end{eqnarray}
The parameters of the model are restricted by the values of the  $W$ and  $Z$  masses  are obtained at tree level  
\begin{eqnarray}
m_W^2 = \frac{g^2}{4}(v_\Phi^2+2v_\Delta^2),\quad m_Z^2 =\frac{g^2}{4\cos^2\theta_W}(v_\Phi^2+4v_\Delta^2), 
\end{eqnarray}
and the electroweak $\rho$ parameter defined  at tree level 
\begin{eqnarray}
\rho \equiv \frac{m_W^2}{m_Z^2\cos^2\theta_W}=\frac{1+\frac{2v_\Delta^2}{v_\Phi^2}}{1+\frac{4v_\Delta^2}{v_\Phi^2}}. \label{rho_triplet}
\end{eqnarray}
As the experimental value of the $\rho$ parameter is near unity, 
$v_\Delta^2/v_\Phi^2$ is required to be much smaller than unity at the tree level, justifying the expansions in Eqs. (\ref{lambda2}), (\ref{lambda3}), (\ref{lambda4}). 
Note that the smallness of $v_\Delta/v_\Phi$ insures that the mixing angles $\beta_\pm$ and $\beta_0$ are close to $0$, while $\alpha$ remains undetermined.
Finally, small Majorana neutrino masses, proportional to the lepton number violating coupling constant $\mu$, are generated by the Yukawa interaction of the triplet field   
\begin{equation}
(m_\nu)_{ij}=\sqrt{2}h_{ij} v_\Delta=h_{ij}\frac{\mu v_\Phi^2}{M_\Delta^2}. \label{mn}
\end{equation}
If $\mu\ll M_\Delta$ the smallness of the neutrino masses are explained by the type II seesaw mechanism. This condition constrains the size of $h_{ij} v_\Delta$ by relating it to the neutrino mass.
The smallness of $v_{\Delta}$ yields approximate relationships among the masses: 
\begin{eqnarray}
&&m_{H^{+}}^2-m_{H^{++}}^2\simeq m_{A}^2-m_{H^+}^2 \simeq\frac{\lambda_5}{4}v_\Phi^2, \label{mass1}\\
&&m_H^2\simeq m_A^2\left(\simeq M_\Delta^2\right), \label{mass2}
\end{eqnarray}
which are valid to ${\cal O}(v_\Delta^2/v_\Phi^2)$.  We can further simplify, for the parameters $\lambda_2 , \lambda_3, \lambda_4$ in terms of  $\lambda_5$,  the neutral Higgs masses and the mixing angle:
\begin{eqnarray}
\lambda_2&=&-\lambda_5+\frac{1}{2v_\Delta^2}\sin^2 \alpha \left(m_h^2-m_H^2\right)+2\frac{m_H^2}{v_\Phi^2} +{\cal O}(\frac{v_\Delta^2}{v_\Phi^2}),\label{lambda2}\\
\lambda_3&=& \lambda_5 + {\cal O}(\frac{v_\Delta^2}{v_\Phi^2}),\label{lambda3}\\
\lambda_4&=&-\lambda_5+ \frac{m_h^2-m_H^2}{2 v_\Delta v_\Phi} \sin 2 \alpha +2\frac{m_H^2}{v_\Phi^2}+{\cal O}(\frac{v_\Delta^2}{v_\Phi^2}) \label{lambda4},
\end{eqnarray}
which must be consistent with conditions on the Higgs potential.
\subsection{Positivity Conditions on the Higgs potential}
\label{sec:lambdas}

The parameters in the Higgs potential are not arbitrary, but subjected to several conditions. These have been thoroughly analyzed in \cite{Arhrib:2011uy}, and we summarize their results briefly. Positivity requirement in the singly and doubly
charged Higgs mass sectors, (for $v_\Delta>0$) are:
\begin{eqnarray}
\mu&>&0;    \qquad 
\mu > \frac{\lambda_5 v_\Delta}{2\sqrt{2}};  \qquad 
\mu > \frac{\lambda_5 v_\Delta}{\sqrt{2}} + \sqrt{2} \frac{ \lambda_3 v_\Delta^3}{v_\Phi^2},  
\end{eqnarray}
 while, for the requirement that the potential is bounded from below, the complete set of conditions are:
 \begin{eqnarray}
&& \lambda_1 > 0 \;\;{;}\;\; \lambda_2+\lambda_3 > 0  \;\;{;}  \;\;\lambda_2+\frac{\lambda_3}{2} > 0;  
\label{eq:potential1} \\
&&  \;\;\lambda_4+ \sqrt{4\lambda_1(\lambda_2+\lambda_3)} > 0 \;\;;\;\;
\lambda_4+ \sqrt{4\lambda_1(\lambda_2+\frac{\lambda_3}{2})} > 0;  \label{eq:potential2}\\
&& 
\;\; \lambda_4+\lambda_5+\sqrt{4\lambda_1(\lambda_2+\lambda_3)} > 0 \;\; {;} \;\; 
\lambda_4+\lambda_5+\sqrt{4\lambda_1(\lambda_2+ \frac{\lambda_3}{2})} > 0.  \label{eq:potential3}
\end{eqnarray}
Note that, from the expressions in Eq. (\ref{eq:lambdavalues}), some conditions are automatically satisfied, such as positivity of $\mu $ and $ \lambda_1$. Positivity of $ \lambda_2+\lambda_3$ and $ \displaystyle  \lambda_2+\frac{\lambda_3}{2}$ are consistent with the requirements of the square root in Eq. (\ref{eq:potential2}) and (\ref{eq:potential3}) being real. 
From all of these conditions, the last expressions in Eq.  (\ref{eq:potential2}) and (\ref{eq:potential3}) would restrict possible enhancements in the $h,H \to \gamma \gamma$ decay. 

Before we proceed with the detailed analysis, some general comments are in order.  
 As shown in \cite{Arhrib:2011vc}, and as we show in detail in the next section, for $\sin \alpha=0$,  the coupling between $h$ and the doubly charged Higgs is strictly proportional to $\lambda_4$. If $\lambda_4$ is positive (negative), the contribution of the doubly-charged Higgs bosons  is subtracted from (added to) the $W$ boson contribution, which is dominant, resulting in a suppression (enhancement) of the $\gamma \gamma $ branching ratio. The contribution for the singly-charged Higgs bosons is significantly smaller, but follows the same general pattern. Note that if $\alpha=0$, $\lambda_2=\lambda_4$. Thus it is inconsistent to assume $\lambda_2>0$, while $\lambda_4<0$. Moreover, for $\sin \alpha=0$,
\begin{equation}
\lambda_4=-\lambda_5+2\frac{m_H^2}{v_\Phi^2}=-2\left (\frac{m_H^2}{v_\Phi^2}-\frac{m_{H^{++}}^2}{v_\Phi^2}\right )+2\frac{m_H^2}{v_\Phi^2}= 2\frac{m_{H^{++}}^2}{v_\Phi^2},
\end{equation}
and thus $\lambda_4$ cannot be negative, preventing an enhancement of $R_{\gamma \gamma}$ for the unmixed neutral Higgs boson $h$ due to the presence of the singly and doubly-charged Higgs in the loop.

Thus the only possibility in which there could be some enhancement in the decay to $\gamma \gamma$ is the case in  which there is some mixing between the two states $\Phi^0$ and $\Delta^0$, and both states are responsible for some of the signals observed at the LHC, Tevatron and LEP, an alternative which we investigate in the reminder of this work. We continue to call the two mixed states $h$ and $H$, with the convention that, when $\alpha \to 0$, these states correspond to the unmixed states $\Phi^0$ (neutral doublet)  and $\Delta^0$ (neutral triplet), respectively. We take a different point of view from previous analyses. We make no assumption about the mixing, but express all parameters as functions of $\alpha$, the mixing angle in the neutral (CP-even) Higgs sector, as in Eqs. (\ref{mass1}), (\ref{mass2}), (\ref{lambda2}), (\ref{lambda3}) and (\ref{lambda4}).

\section{Decay rates of the neutral Higgs bosons in the HTM}
\label{sec:decays}
\subsection{Decay rates to $\gamma \gamma$}
\label{sec:gamgam}

 In this section we present the analytic expressions for the decays of the neutral bosons. The detailed numerical analysis and comparison  with the LHC, Tevatron and LEP data follows in the next section. We concentrate first on the decays to $\gamma \gamma$, as, in spite of the small rate, these decays are very promising, as $M_{\gamma \gamma}$ can be reconstructed to ${\cal O}$(1\%) accuracy. Indeed both CMS and ATLAS have their most accurate data for this channel.  We allow arbitrary mixing in the neutral sector and discuss the restrictions on the parameters in the Higgs sector imposed by the data, as well as by the conditions on the potential, and investigate the consequences for the decay of both neutral Higgs bosons in the HTM.

 First, we return to formulas Eqs.  (\ref{lambda2}), (\ref{lambda3}), (\ref{lambda4}), for $\sin \alpha \ne 0$. As $\lambda_2+\lambda_3>0$, this requires $m_h^2>m_H^2$, that is the state which in the limit $\sin \alpha=0$ is the Higgs doublet state is heavier than the state which in the limit $\sin \alpha=0$ is the Higgs triplet state. Unlike for the state with $\alpha=0$, $\lambda_2 \ne \lambda_4$, more precisely
 \begin{equation}
 \lambda_4=\lambda_2- \frac{m_h^2-m_H^2}{2 v_\Delta^2} \sin \alpha \sin(\alpha-\beta_0)\simeq \lambda_2- \frac{m_h^2-m_H^2}{2 v_\Delta^2} \sin^2 \alpha.
 \label{eq:lam4lam2}
 \end{equation}
 The decay rates of the Higgs bosons in the HTM are defined in terms of the decay of the Higgs boson in the SM (denoted as $\Phi$) as:
\begin{eqnarray}
R_{h,H \to \gamma \gamma}&\equiv& \frac {\sigma_{\rm HTM}(gg \to h, H \to \gamma \gamma)}{\sigma_{\rm SM}(gg \to  \Phi \to \gamma \gamma)} =\frac{[\sigma( gg \to  h, H) \times BR(h,H \to \gamma \gamma)]_{HTM}}{[\sigma( gg \to  \Phi) \times BR(\Phi \to \gamma \gamma )]_{SM}} \nonumber \\
&=&\frac{[\sigma( gg \to  h, H) \times \Gamma(h,H \to \gamma \gamma)]_{HTM}}{[\sigma( gg \to\Phi) \times \Gamma(\Phi \to \gamma \gamma)]_{SM}} \times \frac{[\Gamma( \Phi)]_{SM}}{[\Gamma(h,H)]_{HTM}}, 
\end{eqnarray}
where the ratios of cross section rates by  gluon fusion are:
\begin{equation}
\frac{[\sigma( gg \to  h)]_{HTM}}{[\sigma( gg \to  \Phi)]_{SM}}= \cos^2 \alpha; \qquad \frac{[\sigma( gg \to  H)]_{HTM}}{[\sigma( gg \to  \Phi)]_{SM}}= \sin^2 \alpha.
\end{equation}
We present first at the decay widths of $h$ to $\gamma \gamma$:
\begin{eqnarray}
\label{eq:THM-h2gaga}
[\Gamma(h \rightarrow\gamma\gamma)]_{HTM}
& = & \frac{G_F\alpha^2 m_{h}^3}
{128\sqrt{2}\pi^3} \bigg| \sum_f N^f_c Q_f^2 g_{h ff} 
A_{1/2}
(\tau^h_f) + g_{h WW} A_1 (\tau^h_W) \nonumber \\
&& + \tilde{g}_{h H^\pm\,H^\mp}
A_0(\tau^h_{H^{\pm}})+
 4 \tilde{g}_{h H^{\pm\pm}H^{\mp\mp}}
A_0(\tau^h_{H^{\pm\pm}}) \bigg|^2 \, .
\label{partial_width_htm}
\end{eqnarray}
The couplings of $h$ to the vector bosons and fermions are as follows:
\begin{eqnarray}
g_{h t\overline t}=\cos\alpha/\cos\beta_{\pm} \, 
\label{littleh1tt};  \qquad 
g_{hWW}= \cos\alpha+2\sin\alpha v_\Delta/v_\Phi  \, ,  
\label{littleh1WW} 
\end{eqnarray}
and the scalar trilinear couplings are parametrized as follows:
\begin{eqnarray}
\tilde{g}_{h H^{++}H^{--}}  & = &   \frac{m_W}{ g m_{H^{\pm \pm}}^2} g_{h H^{++}H^{--}} \, ; \label{eq:redgcalhHpp} \quad
\tilde{g}_{h H^+H^-} =  \frac{m_W}{g m_{H^{\pm}}^2} g_{h H^+H^-} \, ,
\label{eq:redgcallittlehHp}
\end{eqnarray}
with the following explicit expressions in terms of the parameters of the scalar potential, Eq. (\ref{pot_htm}):
\begin{equation}
g_{hH^{++}H^{--}} =2\lambda_2v_\Delta \sin\alpha+\lambda_4v_\Phi \cos \alpha \, ,
 \label{eq:ghHpp}
\end{equation}
\begin{eqnarray}
g_{hH^+H^-}=\frac{1}{2}
\bigg\{\left[4v_\Delta(\lambda_2 + \lambda_3) \cos^2{\beta_{\pm}}+2v_\Delta\lambda_4\sin^2{\beta_{\pm}}-
\sqrt{2}\lambda_5v_\Phi  \cos{\beta_{\pm}}\sin{\beta_{\pm}}\right]\sin\alpha  \nonumber \\ 
+\left[4\lambda_1\,v_\Phi  \sin{\beta_{\pm}}^2+{(2\lambda_{4}+\lambda_{5}) }v_\Phi  \cos^2{\beta_{\pm}}+
(4\mu-\sqrt{2}\lambda_5v_\Delta)\cos{\beta_{\pm}}\sin{\beta_{\pm}}\right]\cos\alpha\bigg\} \, .
\label{eq:ghHp}
\end{eqnarray}
These couplings become, in terms of the masses and mixing, 
for the trilinear coupling of the neutral and doubly-charged Higgs to $h$:
 \begin{eqnarray}
 {\tilde g_{hH^{++}H^{--} } }\!\!\!&\simeq &  \frac {v_\Phi^2}{2m_{H^{++}}^2} \left \{ \frac{m_h^2-m_H^2}{v_\Delta v_\Phi} \sin \alpha +\left ( 2 \frac{m_H^2}{v_\Phi^2}-\lambda_5 \right ) \cos (\alpha-\beta_0)  \right \},  \\
  {\tilde g_{hH^{+}H^{-} } }\!\!\!&\simeq &  \frac {v_\Phi^2}{2m_{H^{+}}^2} \left \{ \frac{m_h^2-m_H^2}{v_\Delta v_\Phi} \sin \alpha +\left ( 2 \frac{m_H^2}{v_\Phi^2}-\frac{\lambda_5}{2} \right ) \cos (\alpha-\beta_0)  \right \}.
 \end{eqnarray}
 We obtain similar expressions for the neutral boson $H$.
\begin{eqnarray}
\label{eq:THM-Hgaga}
\Gamma(H \rightarrow\gamma\gamma)
& = & \frac{G_F\alpha^2 m_{H}^3}
{128\sqrt{2}\pi^3} \bigg| \sum_f N_c Q_f^2 g_{H ff} 
A_{1/2}
(\tau^H_f) + g_{H WW} A_1 (\tau^H_W) \nonumber \\
&& + \tilde{g}_{H H^\pm\,H^\mp}
A_0(\tau^H_{H^{\pm}})+
 4 \tilde{g}_{H H^{\pm\pm}H^{\mp\mp}}
A_0(\tau^H_{H^{\pm\pm}}) \bigg|^2 \, .
\end{eqnarray}

The couplings of $H$ to the vector bosons and fermions relative to the values in the SM are as follows:
\begin{eqnarray}
g_{H t\overline t}=-\sin\alpha/\cos\beta_{\pm} \, 
\label{BigH1tt};  \qquad
g_{HWW}= -\sin\alpha+2\cos\alpha v_\Delta/v_\Phi   \, ,  
\label{BigH1WW} 
\end{eqnarray}
The scalar trilinear couplings are parametrized similar to those for $h$:
\begin{eqnarray}
\tilde{g}_{H H^{++}H^{--}}  & = &   \frac{m_W}{ g m_{H^{\pm \pm}}^2} g_{H H^{++}H^{--}} \, ;
 \label{eq:redgcaBiglHHpp} \quad
\tilde{g}_{H H^+H^-} =   \frac{m_W}{g m_{H^{\pm}}^2} g_{H H^+H^-} \, ,
\label{eq:redgcalBigHHp}
\end{eqnarray}
with the following explicit expressions in terms of the parameters of the scalar potential (these can be obtained from the expressions for $h$, with the replacements $\cos \alpha \to -\sin \alpha,~ \sin \alpha \to \cos \alpha$):
\begin{equation}
g_{HH^{++}H^{--}} =2\lambda_2v_\Delta \cos\alpha-\lambda_4v_\Phi  \sin \alpha \, ,
 \label{eq:gHHpp}
\end{equation}
\begin{eqnarray}
g_{HH^+H^-}=\frac{1}{2}
\bigg\{\left[4v_\Delta(\lambda_2 + \lambda_3) \cos^2{\beta_{\pm}}+2v_\Delta\lambda_4\sin^2{\beta_{\pm}}-
\sqrt{2}\lambda_5v_\Phi  \cos{\beta_{\pm}}\sin{\beta_{\pm}}\right]\cos\alpha  \nonumber \\ 
-\left [4\lambda_1\,v_\Phi  \sin{\beta_{\pm}}^2+{(2\lambda_{4}+\lambda_{5}) }v_\Phi  \cos^2{\beta_{\pm}}+
(4\mu-\sqrt{2}\lambda_5v_\Delta)\cos{\beta_{\pm}}\sin{\beta_{\pm}}\right]\sin\alpha\bigg\} \, , \quad
\label{eq:gHHp}
\end{eqnarray}
 we obtain, 
 \begin{eqnarray}
  {\tilde g_{HH^{++}H^{--}}}\!\!\!&\simeq & - \frac{v_\Phi^2}{2m_{H^{++}}^2} \left (2\frac{m_H^2}{v_\Phi^2}-\lambda_5\right) \sin (\alpha-\beta_0) \\
 {\tilde g_{HH^{+}H^{-}}}\!\!\!&\simeq & - \frac{v_\Phi^2}{2m_{H^{+}}^2} \left (2\frac{m_H^2}{v_\Phi^2}-\frac{\lambda_5}{2}\right) \sin (\alpha-\beta_0) 
 \end{eqnarray}
 
We define throughout  $\tau^h_{i}=m^2_{h}/4m^2_{fi}$, $\tau^H_{i}=m^2_{H}/4m^2_{i}$ $(i=f,W,H^{\pm},H^{\pm\pm})$. 
The loop functions $A_1$ (for the $W$ boson) and $A_{1/2}$
(for the fermions, $f$) are  given as
\begin{eqnarray}
A_{0}(\tau) &=& -[\tau -f(\tau)]\, \tau^{-2} \, ,
\label{eq:Ascalar}\\
A_{1/2}(\tau)&=&-\tau^{-1}\left[1+\left(1-\tau^{-1}\right)f\left(\tau^{-1}\right)\right], \label{eq:Afermion}\\
A_1(\tau)&=& 1+\frac32 \tau^{-1}+4 \tau^{-1}\left(1-\frac12 \tau^{-1}\right)f\left(\tau^{-1}\right). \label{eq:Avector} 
\end{eqnarray}
These function are similarly defined for $H$, with the change $h \to H$, and the function $f(\tau)$ is given by :
\begin{eqnarray}
f(\tau)=\left\{
\begin{array}{ll}  \displaystyle
\arcsin^2\sqrt{\tau} & \tau\leq 1 \\
\displaystyle -\frac{1}{4}\left[ \log\frac{1+\sqrt{1-\tau^{-1}}}
{1-\sqrt{1-\tau^{-1}}}-i\pi \right]^2 \hspace{0.5cm} & \tau>1 \, .
\end{array} \right. 
\label{eq:ftau} 
\end{eqnarray}

Second, note that the contribution from the loop with $H^{\pm\pm}$ in Eq.(\ref{partial_width_htm})
is enhanced relative to the contribution from $H^{\pm}$ by a factor of four at the amplitude level. Second, all couplings are evaluated to $\displaystyle{\cal O}(\frac{v_\Delta^2}{v_\Phi^2})$. However, for all relevant parameter space the effect of $\beta_0$ is negligible ($\displaystyle \tan \beta_0= \frac{2v_\Delta}{v_\Phi}$), and we can assume with no loss of generality that $\beta_0\simeq0$. Third, as $m_H$ is the lightest of the two Higgs states (and we would wish to associate it with one of the observed bosons), $\lambda_5$ is constrained to be negative, otherwise the singly and doubly charged Higgs bosons would be unacceptably light.   Fourth, inspection of the analytic expressions indicate that for all of the parameter space, the reduced couplings ${\tilde g}_{hH^{++}H^{--}} $  and ${\tilde g}_{hH^{++}H^{--}} $  are positive (as $\alpha \in (0, \pi/2)$), while ${\tilde g}_{HH^{++}H^{--}} $  and ${\tilde g}_{HH^{++}H^{--}} $ are negative. This means that we expect that, from trilinear couplings alone,  $R(h \to \gamma \gamma)$ could be enhanced with respect to the SM over a region of the parameter space, while $R(H \to \gamma \gamma)$ will be suppressed over all of the parameter space.

\subsection{Branching ratios enhancement of due to  Higgs widths in HTM}
\label{sec:width}

Our considerations for relative branching ratios are affected by the fact the total width of the Higgs boson in the HTM is not the same as in the SM. The widths are the same as those in the SM for  $h$ in the limit $\sin \alpha \to 0$. However, for $\alpha \ne 0$ we must take into account the relative widths factors
\begin{eqnarray}
\frac{[\Gamma(\Phi)]_{SM}}{[\Gamma(h,H)]_{HTM}}=\frac{[\Gamma (\Phi \to \sum\limits_f f {\bar f})+ \Gamma ( \Phi \to WW^*) +\Gamma ( \Phi \to ZZ^*)]_{SM}}{[\Gamma (h, H \to \sum\limits_f f {\bar f})+ \Gamma (h, H \to WW^*) +\Gamma (h, H \to ZZ^*)+ \Gamma (h, H \to \nu \nu)]_{HTM}} \ . \nonumber \\
\end{eqnarray}
We expect this to enhance the relative signal strength, as roughly
$$ \frac{[\Gamma(\Phi)]_{SM}}{[\Gamma(h)]_{HTM}} \simeq \frac{1}{\cos^2 \alpha}\left [ 1-\frac{[\Gamma(h\to \nu \nu)]_{HTM}}{[\Gamma( \Phi)]_{SM}} \right];  \quad  \frac{[\Gamma(\Phi)]_{SM}}{[\Gamma(H)]_{HTM}} \simeq \frac{1}{\sin^2 \alpha}\left [ 1-\frac{[\Gamma(H \to \nu \nu)]_{HTM}}{[\Gamma( \Phi)]_{SM}} \right] \ .$$
In the detailed numerical analysis, we highlight the relative width enhancement to illustrate its importance.

\subsection{Tree-level decays of the Higgs bosons into fermions and gauge bosons}
\label{sec:tree}

The largest branching ratio of a Higgs boson with mass 125 GeV would be to $b \bar b$.  Unfortunately, this channel is very difficult to observe at the LHC as the continuum background exceeds the signal by roughly eight orders of magnitude. The decay into $\tau^+ \tau^-$ is also problematic, because of the low velocity of the Higgs boson, which makes the reconstruction of $m_{\tau \tau}$ difficult. Although observation of  the decays to fermions is problematic, more statistics and combining LHC and Tevatron results will improve data. Thus we include the predictions of the model here.

The decays to the gauge bosons are more promising, but there are also some issues which need to be resolved in interpreting the data there. The decay to $W^\pm W^\mp$ has a large rate, but once one of the $W$ bosons decay leptonically, the Higgs mass is hard to reconstruct, and the analysis relies on angular correlations. The two $W$ bosons are produced with opposite polarization, and as $W$ bosons are purely left-handed the two leptons prefer to move in the same, rather than in opposite directions. On the positive side, the backgrounds are electroweak, and thus small.
The Higgs decay into $ZZ$, with the further decay into four muons,  is referred to as the ``golden channel". This is because the $m_{4l}$ is easy to reconstruct. The limitations are the leptonic branching ratio of the $Z$, and sharp drop in the off-shell Higgs branching ratio.

We show, for completeness, the relative decays branching ratios of the neutral bosons $h$ and $H$ into fermions, as well as into gauge bosons, compared to the SM ones.
The decay rates for $h$ can be expressed as 
\begin{eqnarray}
\Gamma(h\to f\bar{f})&=&\sqrt{2}G_F\frac{m_hm_f^2}{8\pi}N_c^f\beta\left(\frac{m_f^2}{m_h^2}\right)^3\cos^2\alpha,\\
\Gamma(h \to \nu \nu)&=&\Gamma(h \to \nu^c \bar{\nu})+\Gamma(h \to \bar{\nu}^c \nu)
=\sum_{i,j=1}^3S_{ij}|h_{ij}|^2\frac{m_h}{4\pi}\sin^2\alpha,
\end{eqnarray}
The second decay is of the form $h \to$ invisible, as it shows only as missing energy. It does not exist for a SM Higgs, and it is not a good signature for detection at the LHC. Fortunately, this decay width is small, even for $\sin \alpha=1$, as the couplings $h_{ij}$ must be small to generate small neutrino masses. But as these decays are tree-level, we include them in the total width consideration. 

The decay rate of the Higgs boson $h$ decaying into the gauge boson  pair $VV$ ($V=W$ or $Z$) is given by 
\begin{eqnarray}
\Gamma(h\to VV)&=&\frac{ |\kappa_V(h)|^2m_h^3}{128 \pi m_V^4}    \delta_V\left[1-\frac{4m_V^2}{m_h^2}+\frac{12m_V^4}{m_h^4}\right]\beta\left(\frac{m_V^2}{m_h^2}\right), 
\end{eqnarray}
where $\delta_W=2$ and $\delta_Z=1$, and where $\kappa_V(h)$ are the couplings of the Higgs $h$ with the vector bosons:
\begin{eqnarray}
\kappa_W(h)&=&\frac{ig^2}{2} \left( v_\phi\cos \alpha+ 2 v_\Delta \sin \alpha \right), \\
\kappa_Z(h)&=&\frac{ig^2}{2 \cos^2\theta_W} \left( v_\phi\cos \alpha+ 4 v_\Delta \sin \alpha \right).
\end{eqnarray}
The decay rates of the three body decay modes are, 
\begin{eqnarray}
\Gamma(h\to VV^*)&=&\frac{3g_V^2 |\kappa_V(h)|^2m_h}{512 \pi^3 m_V^2}\delta_{V'}F(\frac{m_V^2}{m_h^2}),\end{eqnarray}
where $\delta_{W'}=1$ and $\displaystyle \delta_{Z'}=\frac{7}{12}-\frac{10}{9} \sin^2 \theta_W +\frac{40}{27} \sin^4 \theta_W$, and  
where the function $F(x)$ is given as 
\begin{eqnarray}
F(x)&=&-|1-x|\left(\frac{47}{2}x-\frac{13}{2}+\frac{1}{x}\right)+3(1-6x+4x^2)|\log \sqrt{x}|\nonumber\\
&+&\frac{3(1-8x+20x^2)}{\sqrt{4x-1}}\arccos \left(\frac{3x-1}{2x^{3/2}}\right).\label{decay_F}
\end{eqnarray}

The decay rates for $H$ can be expressed as 
\begin{eqnarray}
\Gamma(H\to f\bar{f})&=&\sqrt{2}G_F\frac{m_Hm_f^2}{8\pi}N_c^f\beta\left(\frac{m_f^2}{m_H^2}\right)^3\sin^2\alpha,\\
\Gamma(H \to \nu \nu)&=&\Gamma(H \to \nu^c \bar{\nu})+\Gamma(H \to \bar{\nu}^c \nu)
=\sum_{i,j=1}^3S_{ij}|h_{ij}|^2\frac{m_H}{4\pi}\cos^2\alpha,~~~~~~~
\end{eqnarray}
with the second expression for the decay $H \to$ invisible.

As before, we can write general formulas for the decay rates of the Higgs boson decaying into the gauge boson $V$ pair  ($V=W$ or $Z$) are given by 
\begin{eqnarray}
\Gamma(H\to VV)&=&\frac{|\kappa_V(H)|^2m_H^3}{128 \pi m_V^4}    \delta_V\left[1-\frac{4m_V^2}{m_H^2}+\frac{12m_V^4}{m_H^4}\right]\beta\left(\frac{m_V^2}{m_H^2}\right),
\end{eqnarray}
where $\delta_W=2$ and $\delta_Z=1$ and where $\kappa_V(H)$ are the couplings of the Higgs $H$ with the vector bosons:
\begin{eqnarray}
\kappa_W(H)&=&\frac{ig^2}{2} \left( -v_\phi\sin \alpha+ 2 v_\Delta \cos \alpha \right), \\
\kappa_Z(H)&=&\frac{ig^2}{2 \cos^2\theta_W} \left( -v_\phi\sin \alpha+ 4 v_\Delta \cos \alpha \right).
\end{eqnarray}
The decay rates of the three body decay modes are, 
\begin{eqnarray}
\Gamma(H\to VV^*)&=&\frac{3g_V^2|\kappa_V(H)|^2} {512 \pi^3 m_V^2}m_H\delta_{V'}F(\frac{m_V^2}{m_H^2}),
\end{eqnarray}
where $\delta_{W'}=1$ and $\displaystyle \delta_{Z'}=\frac{7}{12}-\frac{10}{9} \sin^2 \theta_W +\frac{40}{27} \sin^4 \theta_W$, and  
where
the function $F(x)$ is given in Eq. (\ref{decay_F}).

\section{Analysis of the decays of $h$ and $H$ }
\label{sec:analysis}

We proceed by evaluating the branching ratios into photons of both $h$ and $H$ in three scenarios, motivated by existing data. We summarize the experimental constraints for the  state at $125$ GeV in Table \ref{tab}, and list the additional properties of the Higgs bosons specific to each Scenario.
\begin{itemize}
\item Scenario 1 (the LHC/CMS Scenario): $m_H=125$ GeV, $m_h=136$ GeV.
In this scenario we require, for the state $h$ at 136 GeV, in addition to the conditions in Table \ref{tab}, that   $R(h \to \gamma \gamma)=0.45 \pm 0.3,~ R(h \to ZZ^*) \le 0.2$, and $R(h \to \tau \tau)<1.8$, in agreement with the excess observed by CMS. 
\item Scenario 2 (the LEP/LHC Scenario): $m_H=98$ GeV, $m_h=125$ GeV.
In this scenario we require $0.1< R(H \to b{\bar b})<0.25$ in agreement with the excess in $e^+e^-$ at LEP, and for $h$, the  conditions from Table \ref{tab}.
\item Scenario 3 (almost degenerate ATLAS and CMS Scenario): the two CP-even neutral Higgs bosons $H$ and $h$ are  (almost) degenerate and have both mass of about $125$ GeV. In this case, we sum over the relative width $R(h)$ and $R(H)$ of both bosons and compare with the signal at $125$ GeV with the conditions from Table \ref{tab}.
\end{itemize}
Additionally, we also comment on the case in which one of the Higgs states is the one seen at the LHC at 125 GeV, and the other has escaped detection. Throughout the analysis, we impose no restrictions on the mixing and  express all the masses and couplings as a function of $\sin \alpha$ and the mass splitting parameter $\lambda_5$.

\begin{table}
\begin{center}
\begin{tabular}{c|c|c|c|c|c}
\hline
\hline
$Experiment$  & $R^{exp}_{\gamma\gamma}$ & $R^{exp}_{ZZ^*}$ & $R^{exp}_{WW^*}$ & $R^{exp}_{bb}$ & $R^{exp}_{\tau\tau}$ \\
\hline
(1) CMS 7+8 TeV & $1.56\pm 0.43$ & $0.7\pm 0.44$ & $0.6\pm 0.4$ & $0.12\pm 0.70$ & $-0.18\pm 0.75$ \\
\hline
(2) ATLAS 7+8 TeV & $1.9\pm 0.5$ & $1.3\pm 0.6$ & -- & -- & -- \\
\hline
(3) CDF and D0 & $3.6\pm 2.76$ & -- & $0.32\pm 0.83$ & $1.97\pm 0.71$ & --\\
\hline
\hline
\end{tabular}
\end{center}
\caption{Experimental data from LHC and the Tevatron  for the boson at 125 GeV.}
\label{tab}
\end{table}

\subsection{$\gamma\gamma$ decays for mixed neutral Higgs}

\subsubsection{Scenario 1}

\begin{figure}[t]
\center
\vskip -0.3in 
\begin{center}
$
	\begin{array}{cc}
\hspace*{-0.2cm}
	\includegraphics[width=3.2in,height=3.0in]{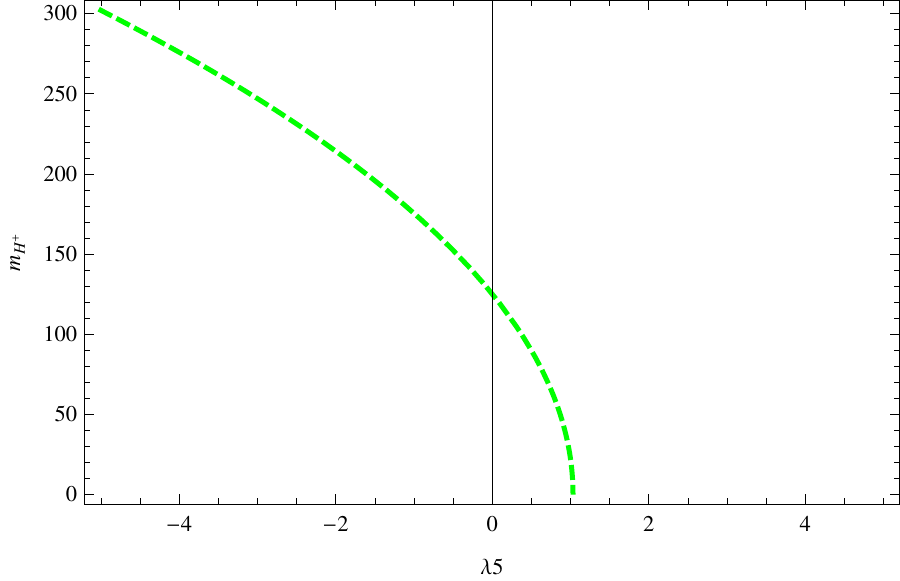}
&
	\includegraphics[width=3.2in,height=3.0in]{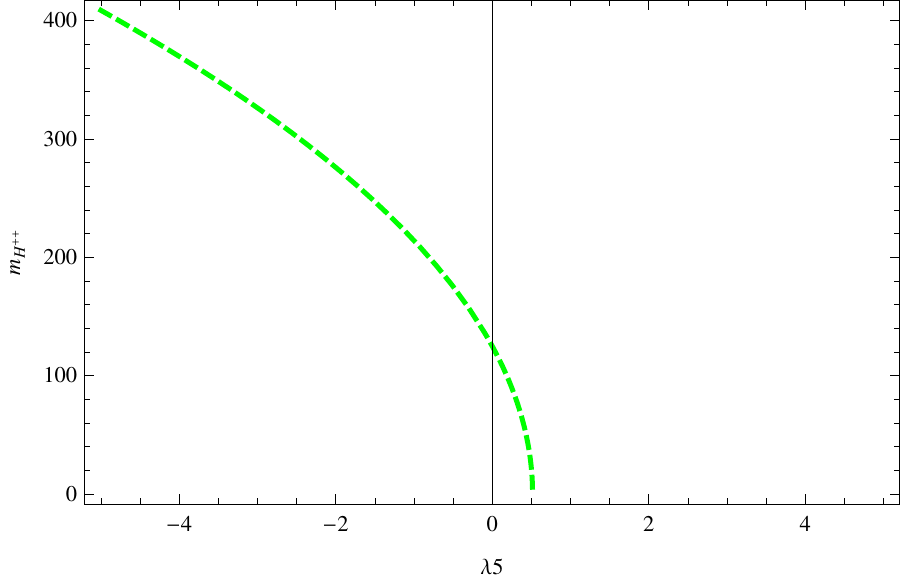}
        \end{array}$
        \end{center}
\caption{Values of the singly charged (left panel) and doubly charged Higgs masses (right panel)  in Scenario 1, with the parameter $\lambda_5\approx \frac{4}{v_\Phi^2}(m_H^2-m_{H^+}^2) \approx \frac{4}{v_\Phi^2}(m_{H^+}^2-m_{H^{++}}^2)$.} 
\label{mhp1}
\end{figure}

We study the implications on the parameter space of the HTM if the lightest Higgs boson is the one observed at the LHC, with the mild $\gamma \gamma$ excess at CMS being due to a second Higgs boson at 136 GeV. Setting these values for the $h$ and $H$ masses, we plot the masses of  the singly and doubly charged Higgs in Fig. \ref{mhp1} as functions of $\lambda_5$. The graphs justify our expectations, based on analytical results, that $\lambda_5$ must be negative, yielding the ordering $m_{H^{++}}>m_{H^+}>m_H$. These graphs also give the values of the charged Higgs masses for different $\lambda_5$ values, to be used in the explorations of $R(\gamma \gamma)$. As $\lambda_4+\lambda_5+\sqrt{4\lambda_1(\lambda_2+ \frac{\lambda_3}{2})} > 0$, we checked the relationship between the $\lambda$'s for $\lambda_5<0$, and found that the inequality is satisfied over the whole parameter space. 
Before proceeding with the analysis, we note that  we are dealing with a very different parameter space than for $\alpha=0$. The  states are now mixed significantly, the state $H$ which in the limit $\alpha \to 0$ is neutral triplet Higgs boson is lighter than the  state $h$ which in the limit $\alpha \to 0$ is neutral doublet Higgs boson, and the ordering of mass states is opposite to that favored for $\alpha=0$ \cite{Aoki:2011pz}, that is in our model $m_{H^{++}}>m_{H^+}>m_H$. 

In order to proceed with the analysis of Higgs decays, we must set reasonable, but not over-conservative limits on doubly-charged boson masses. The strongest limits on the doubly-charged boson masses come from ATLAS \cite{ATLAS:2012hi} and CMS  \cite{Chatrchyan:2012ya}, from $pp \to H^{\pm \pm}H^{\mp \mp}$. At ATLAS, assuming a branching ratio of 100\% into left-handed leptonic final states, masses of  less than 409 GeV, 375 GeV and 398 GeV are excluded, for Yukawa couplings of $h_{ij}>3 \times 10^{-6}$, for final states $e^\pm e^\pm$, $e^\pm \mu^\pm$ and $\mu^\pm \mu^\pm$ respectively. These confirm, and are slightly more stringent than the Tevatron measurements \cite{Abazov:2004au,:2008iy,Acosta:2004uj}. Separate searches were performed  for $q{\bar q} \to \gamma^*, Z^* \to H^{\pm \pm}H^{\mp \mp}$ and $q'{\bar q} \to W^* \to H^{\pm \pm}H^{\mp \mp}$.  For cases where the final state has one or two $\tau^\pm$ leptons, the limits are weaker, 350 GeV and 200 GeV, respectively \cite{Aaltonen:2008ip}. However, most of these limits have been obtained  for  complete dominance of the leptonic decays (which is the case for $v_{\Delta}<10^{-4}$ GeV), and degeneracy of the triplet scalars. In this work, we assume $v_\Delta \sim {\cal O}(1$ GeV), for which the $H^{\pm \pm} \to W^\pm W^\pm$ dominates \cite{Akeroyd:2009hb}. The scenario in which the doubly-charged Higgs decay predominantly into two same-sign vector bosons  has been explored, and it was shown that the LHC running at 8 or 14 TeV would be able to detect such a boson with a mass of $\sim$ 180 GeV. Additionally, for the case where $m_{H^{\pm \pm}} >m_{H^\pm}$, as it in this case, the decay $H^{\pm\pm} \to H^\pm W^{\pm\,*}$ can be dominant over a large range of $v_\Delta$ \cite{Akeroyd:2005gt}. In view of all these considerations, we wish to keep our analysis as general as possible so 
 we consider $m_{H^\pm}$ as low as 110 GeV, and $m_{H^{\pm \pm}}$ as low as 150 GeV. From Fig, \ref{mhp1} this requires that $\lambda_5$ is negative, and from the figure, if $|\lambda_5|>1/2$, $m_{H^{++}}>175$ GeV and  $m_{H^{+}}>150$ GeV.

\begin{figure}[t]
\center
\begin{center}
$
	\begin{array}{cc}
\hspace*{-0.6cm}
\includegraphics[width=3.8in,height=3.8in]{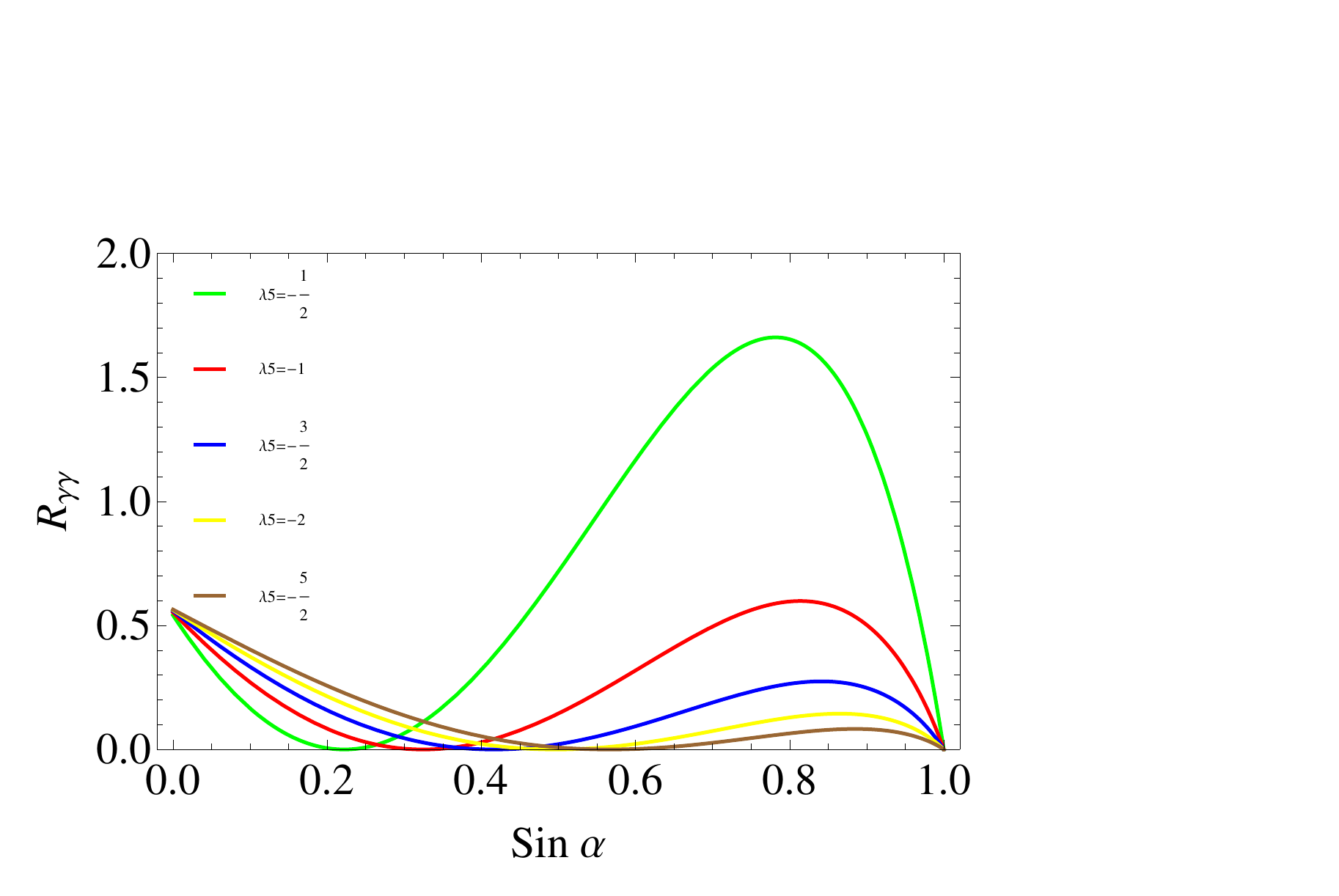}
&\hspace*{-2.0cm}
	\includegraphics[width=3.8in,height=3.8in]{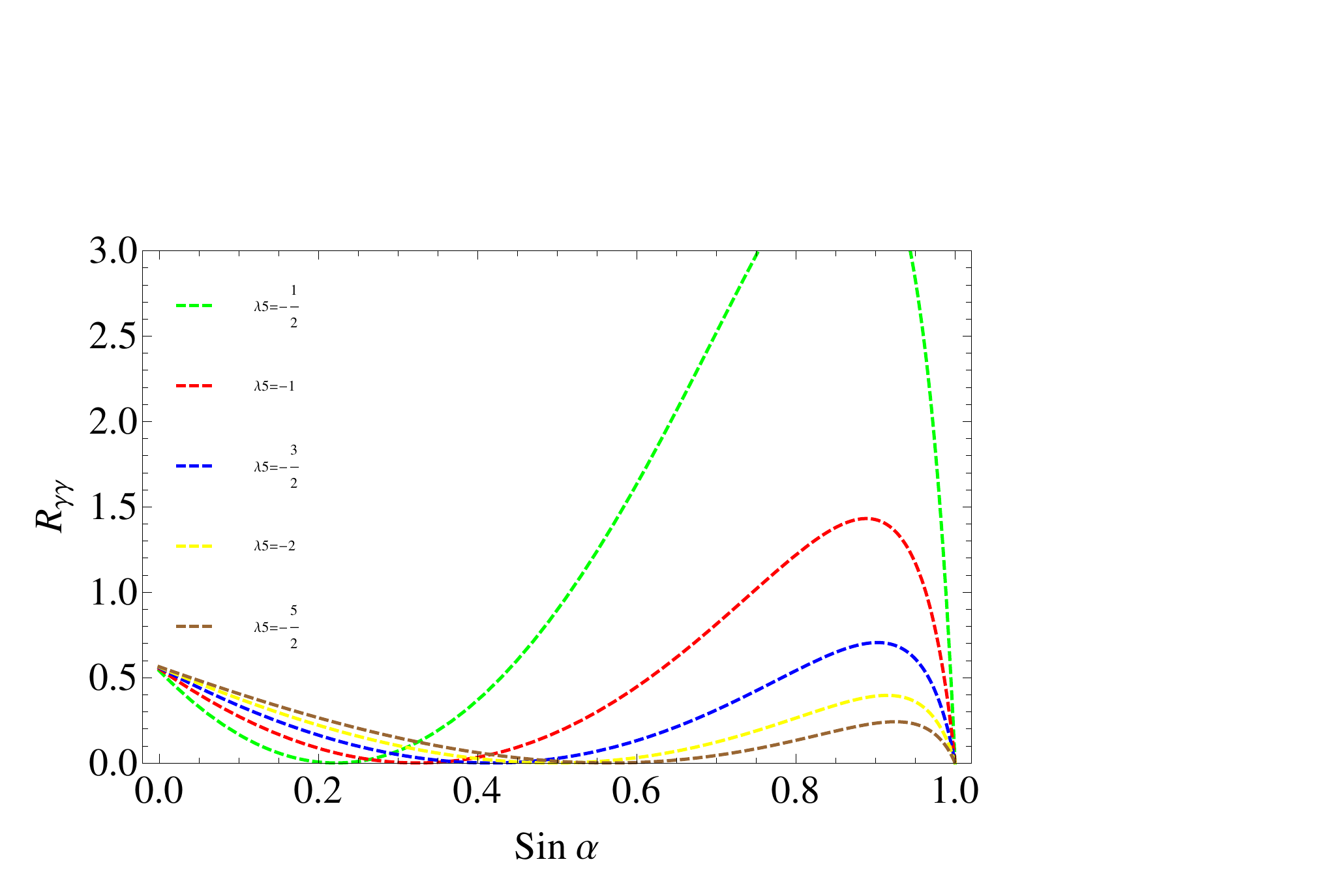}\\
	\hspace*{-0.6cm}
	\includegraphics[width=3.8in, height=3.8in]{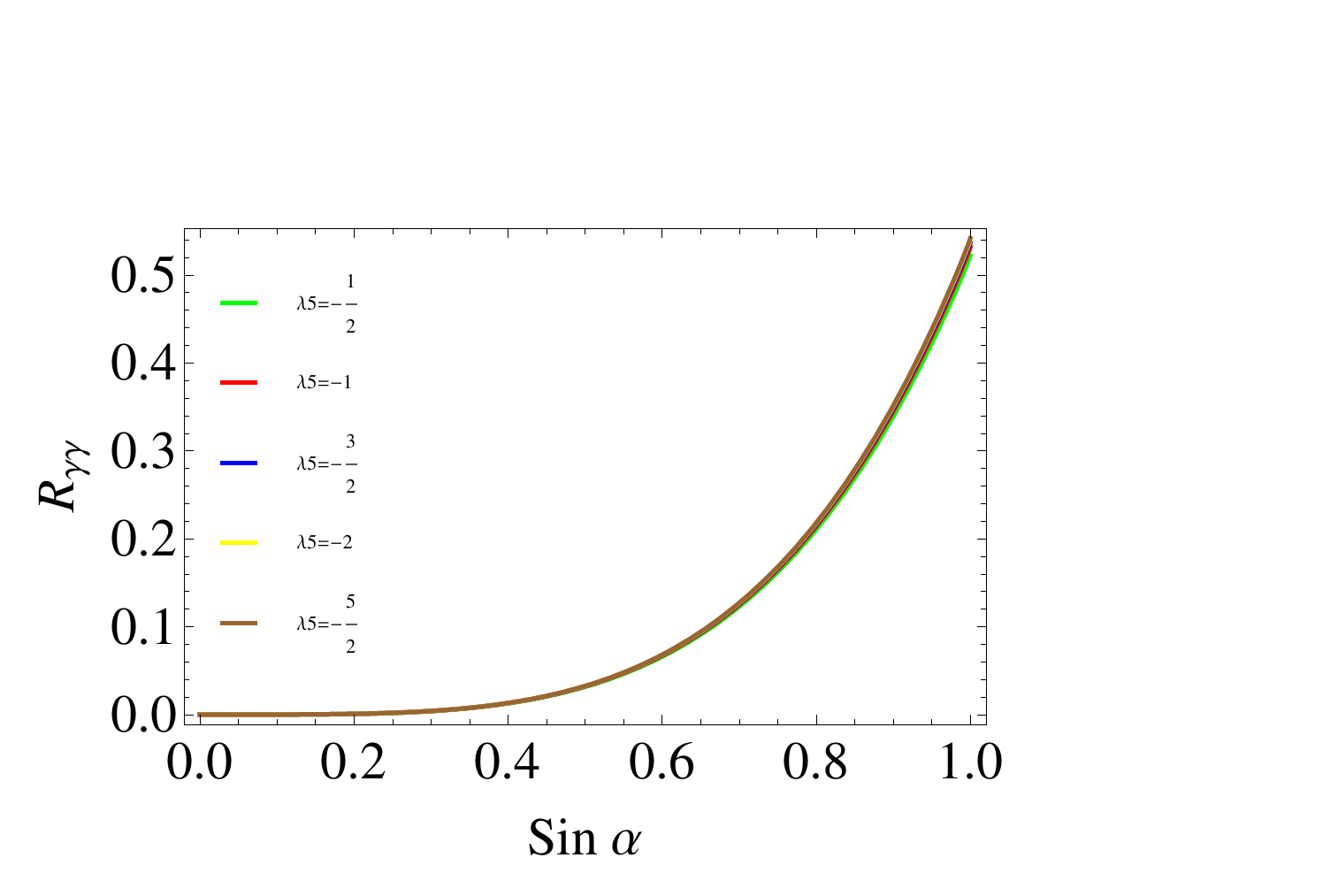}
&\hspace*{-2.0cm}
	\includegraphics[width=3.8in,height=3.8in]{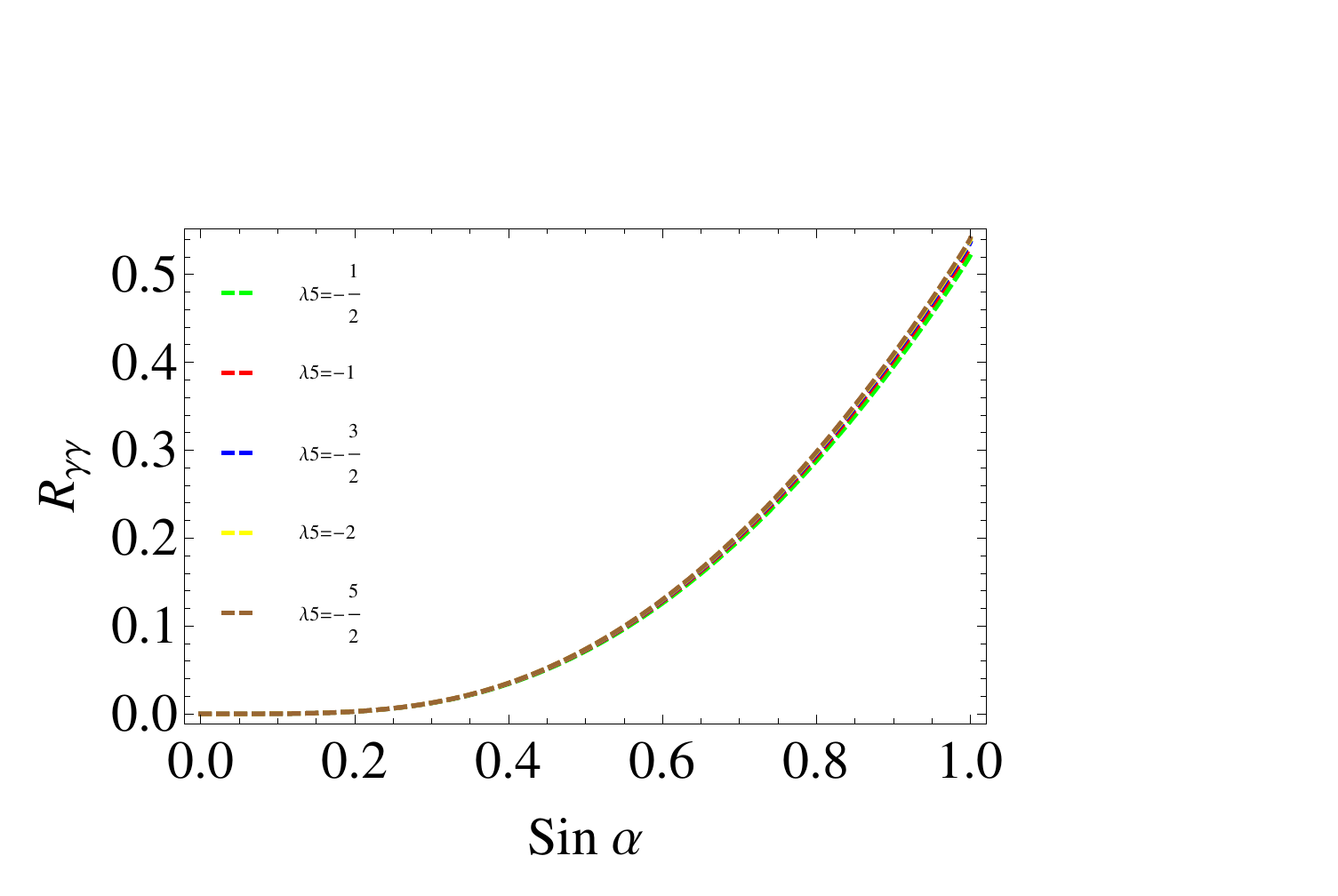}
        \end{array}$
        \end{center}
\caption{Decay rates for $h \to \gamma \gamma$  (top row) and $H \to \gamma \gamma$  (bottom row) as a function of $\sin \alpha$ in Scenario 1, for different values of the parameter $\lambda_5$. The left-handed panels show the relative widths uncorrected for relative width differences, the right-handed panels include the total width corrections. } 
\label{1gamgam}
\end{figure}

We present next the plots for the relative signal strength (with respect to the SM one) of $R_{h \to \gamma \gamma}$ and $R_{H \to \gamma \gamma}$ as a function of $\sin \alpha$, for various values of $\lambda_5$. For each value of $\lambda_5$, we obtain $m_{H^{+}}$ and $m_{H^{++}}$, and introduce these values into calculation of the branching ratios. The plots are in Fig. \ref{1gamgam}, top for $h$, bottom for $H$, and at the left, without width corrections, at the right, including width corrections. Increasing the absolute value of $\lambda_5$ increases the charged Higgs masses and depresses the relative ratio of decay into $\gamma \gamma$.  While the decay of the heavier Higgs boson (at 136 GeV) can be enhanced significantly or suppressed with respect to the SM, fulfilling  the constraint $R(h \to \gamma \gamma)=0.45 \pm 0.3$ for several $\lambda_5$ values, the lighter boson signal is always reduced with respect to the SM. Thus, $H$ cannot be the boson observed at the LHC with mass of 125 GeV, confirming  our analytical considerations, and this scenario is disfavored by the present LHC data.

%

\subsubsection {Scenario 2}

\begin{figure}[t]
\center
\vskip -0.3in 
\begin{center}
$
	\begin{array}{cc}
\hspace*{-0.2cm}
	\includegraphics[width=3.2in,height=3.0in]{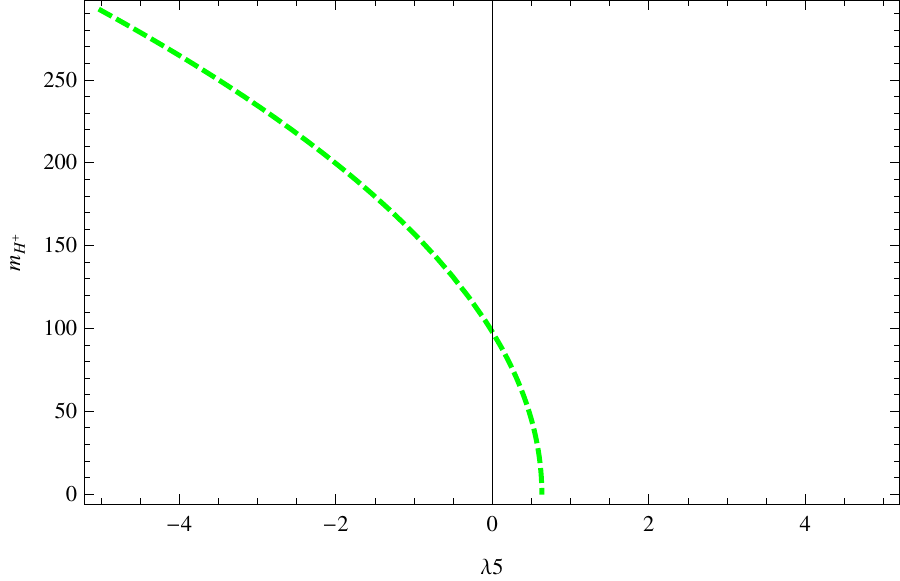}
&
	\includegraphics[width=3.2in,height=3.0in]{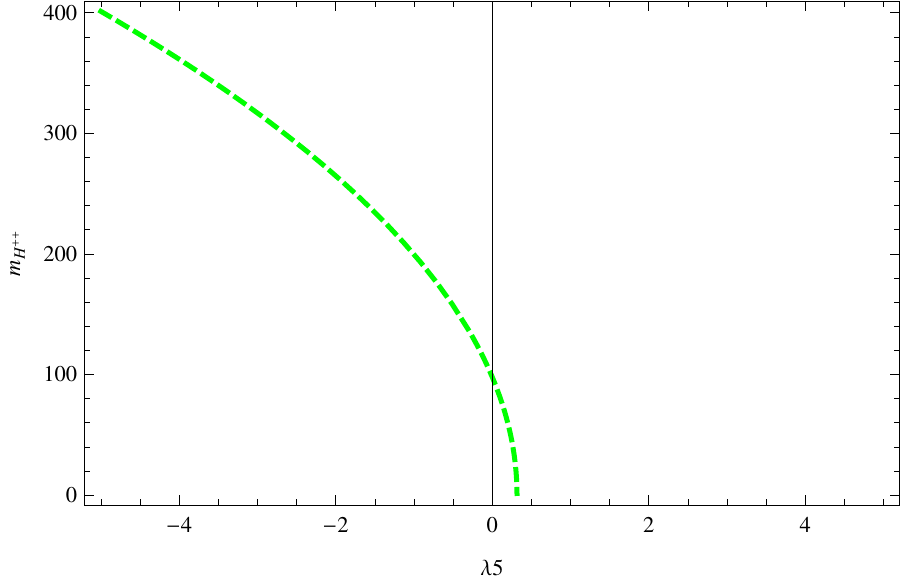}
        \end{array}$
        \end{center}
 \hspace*{-10cm}
\caption{Values of the singly charged (left panel) and doubly charged Higgs masses (right panel)  in Scenario 2, with the parameter $\lambda_5=\frac{4}{v_\Phi^2}(m_H^2-m_{H^+}^2) \approx \frac{4}{v_\Phi^2}(m_{H^+}^2-m_{H^{++}}^2)$.} 
\label{mhp2}
\end{figure}

We now proceed to analyze the implications on the parameter space of the HTM if the lightest Higgs boson is the $2.3\sigma$ signal excess observed at  LEP at 98 GeV, while the heavier Higgs boson is the  boson observed at the LHC at 125 GeV. Setting these values for the $h$ and $H$ masses, we plot the masses of  the singly and doubly charged Higgs in Fig. \ref{mhp2} as functions of $\lambda_5$ (singly charged at the left, doubly charged at the right). Again, in this scenario  $\lambda_5$ is constrained to be  negative, yielding the ordering $m_{H^{++}}>m_{H^+}>m_H$. From the figure  if $|\lambda_5|>1/2$, $m_{H^{++}}>160$ GeV and  $m_{H^{+}}>130$ GeV.
The values of the charged Higgs masses for different $\lambda_5$ values, shown in Fig. \ref{mhp2} are then used in the explorations of $R(\gamma \gamma)$.

\begin{figure}[t]
\center
\begin{center}
$
	\begin{array}{cc}
\hspace*{-0.6cm}
\includegraphics[width=3.8in,height=3.8in]{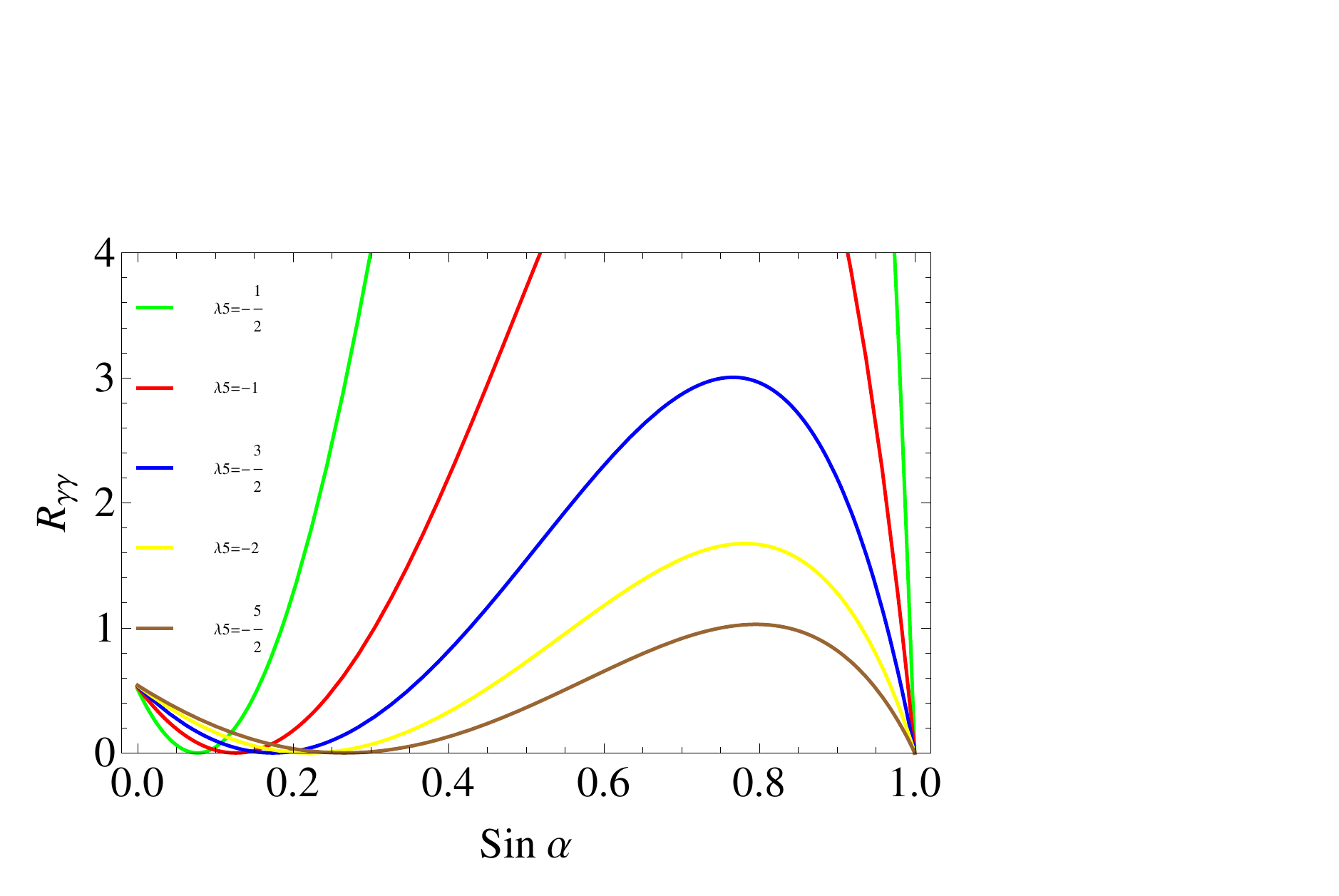}
&\hspace*{-1.0cm}
	\includegraphics[width=3.8in,height=3.8in]{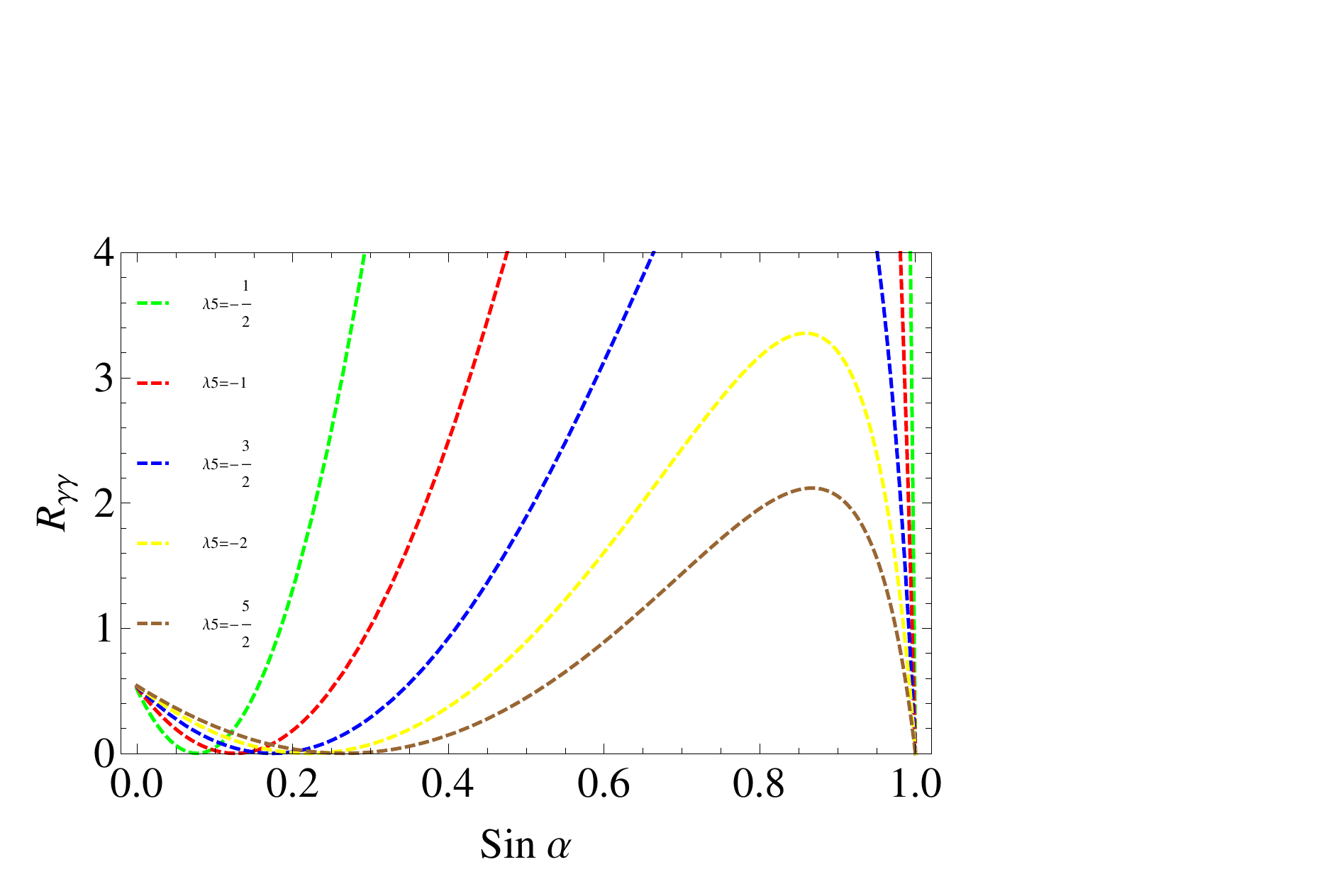}\\
	\hspace*{-0.6cm}
	\includegraphics[width=3.8in, height=3.8in]{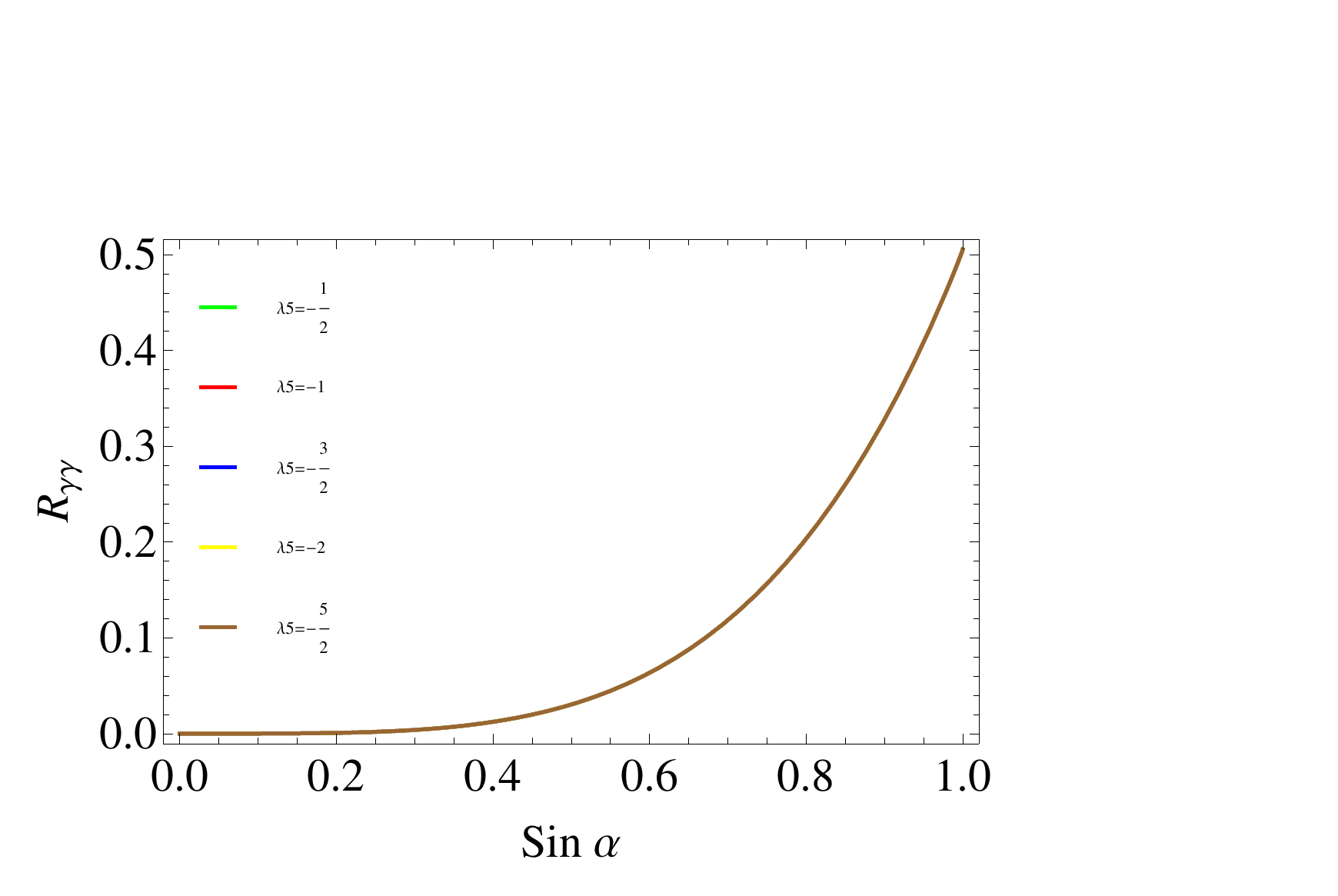}
&\hspace*{-2.0cm}
	\includegraphics[width=3.8in,height=3.8in]{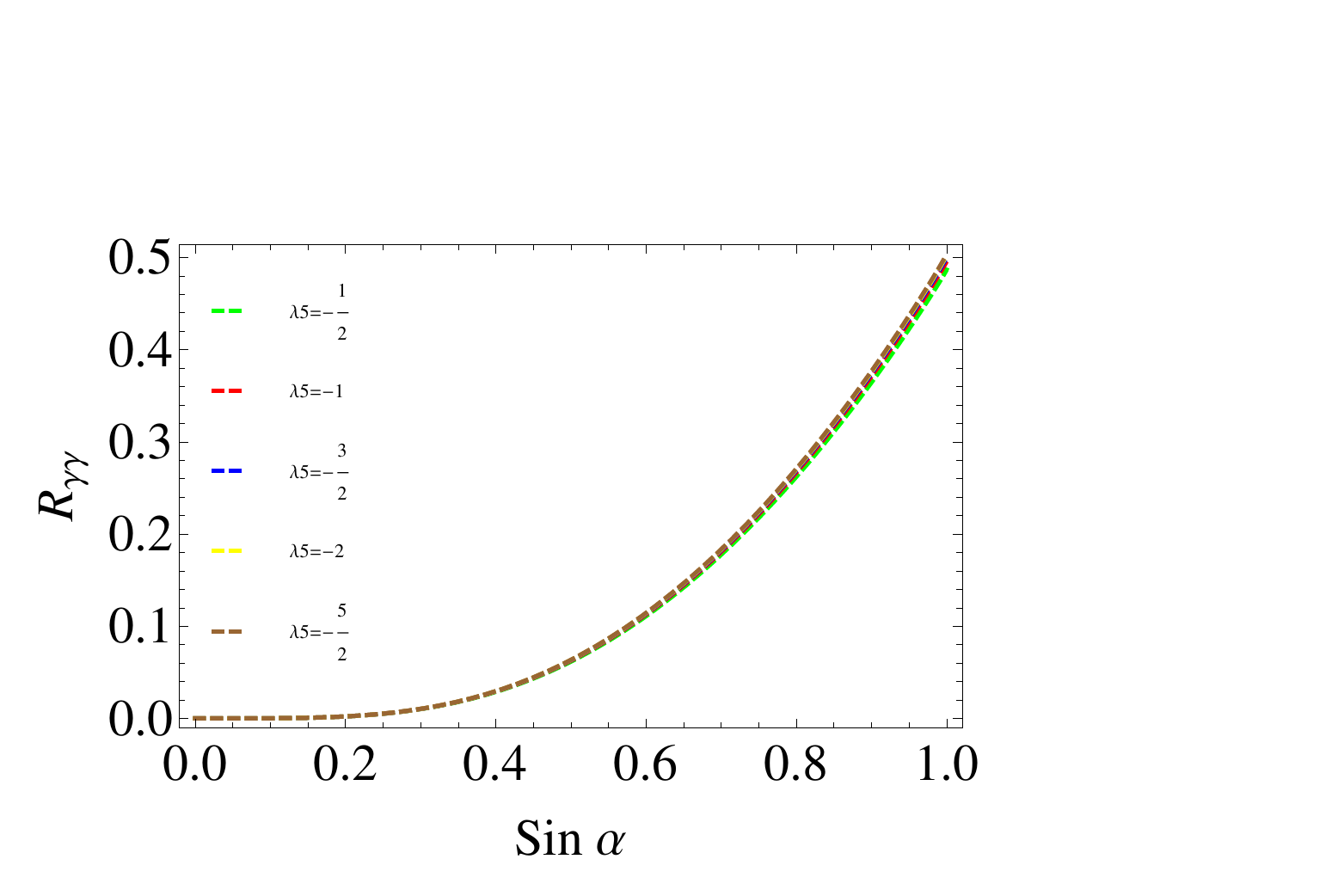}
        \end{array}$
        \end{center}
\caption{Decay rates for $h \to \gamma \gamma$  (top row) and $H \to \gamma \gamma$  (bottom row) as a function of $\sin \alpha$ in Scenario 2, for different values of the parameter $\lambda_5$. The left-handed panels show the relative widths uncorrected for relative width differences, the right-handed panels include the total width corrections.} 
\label{2gamgam}
\end{figure}

We present the plots for the relative signal strength (with respect to the SM one) of $R_{h \to \gamma \gamma}$ and $R_{H \to \gamma \gamma}$ as a function of $\sin \alpha$, for various values of $\lambda_5$ in Fig. \ref{2gamgam}, on the top row for $h$, and the bottom one for $H$. The left panels show the relative $\gamma \gamma$ widths uncorrected for relative width differences and the right-handed panels include the total width corrections. While the decay of the heavier Higgs boson (at 125 GeV) can be enhanced significantly with respect to the SM, the lighter boson signal is always reduced with respect to the SM. If the charged Higgs bosons are relatively light, the angle for which the enhancement is about a factor of  1.5-2  times the SM value is about $\sin \alpha \simeq 0.2$ for $\lambda_5=-1/2$, about $\sin \alpha \simeq 0.35$ for $\lambda_5=-1$, about $\sin \alpha \simeq 0.5$ for $\lambda_5=-3/2$,  and in a  range $\sin \alpha \simeq 0.6-0.9 $ for $\lambda_5=-2$. For the latter case, $m_{H^{++}}=260$ GeV and  $m_{H^{+}}=200$ GeV. For all of the parameter ranges where $h \to \gamma \gamma$ is enhanced, the width of the other neutral Higgs boson $H \to \gamma \gamma $ is suppressed and thus this Higgs boson would escape detection. This feature is general: as long as $h$ is the boson observed at 125 GeV, and $H$ lies below, the decay to $H \to \gamma \gamma$ would be suppressed with respect to a SM Higgs boson of the same mass, and $H$ would escape detection. Thus this scenario would survive even if the LEP $e^+e^-$ excess at 98 GeV does not. The details of the exact enhancements depend on the mass splittings, but the enhancements of $h \to \gamma \gamma$ themselves appear to be fairly robust.

\subsubsection {Scenario 3}

\begin{figure}[t]
\center
\begin{center}
$
	\begin{array}{cc}
\hspace*{-0.6cm}
\includegraphics[width=3.8in,height=3.8in]{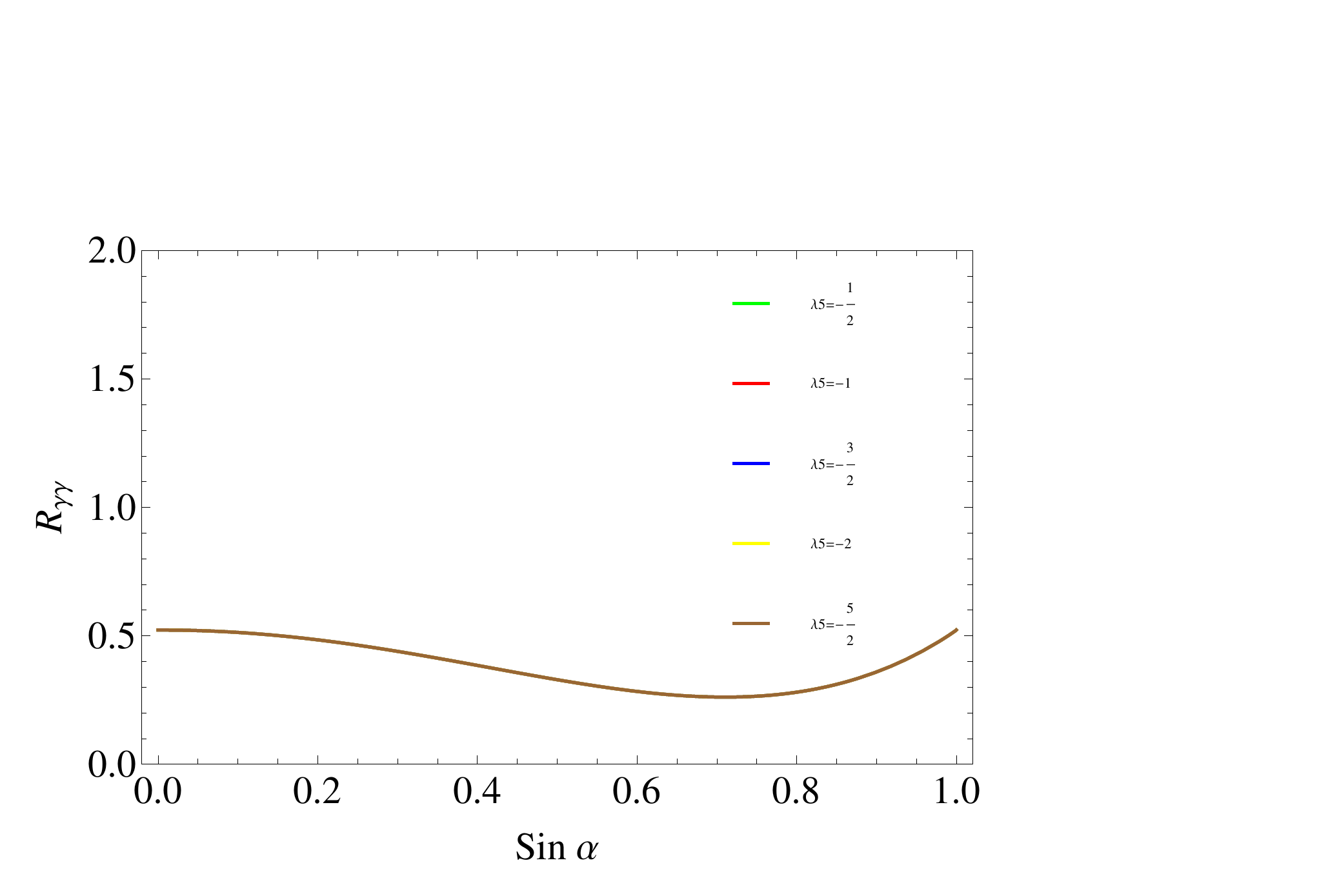}
&\hspace*{-1.0cm}
	\includegraphics[width=3.8in,height=3.8in]{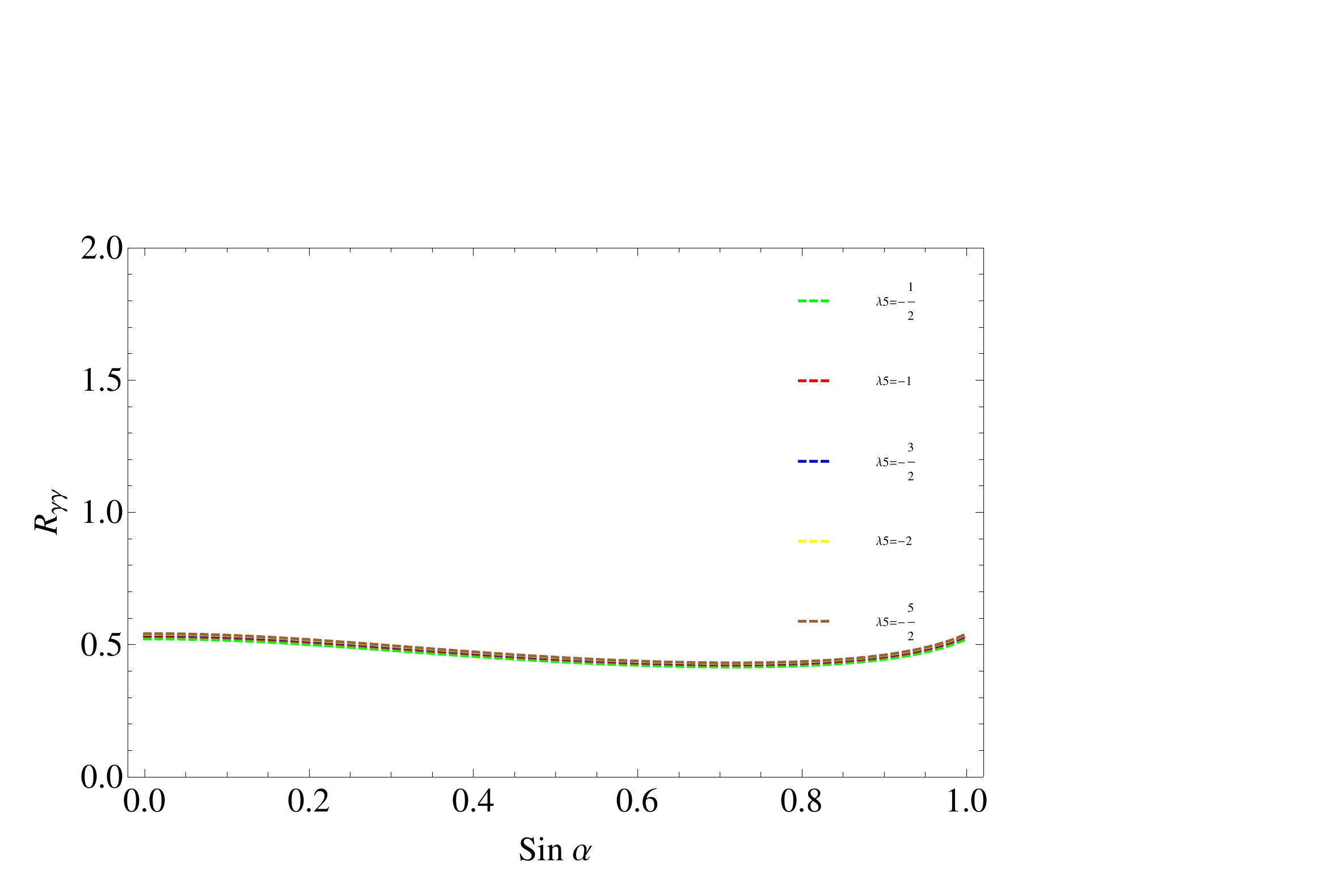}
        \end{array}$
        \end{center}
\caption{Decay rates for $h/H \to \gamma \gamma$   as a function of $\sin \alpha$ , for different values of the parameter $\lambda_5$ in Scenario 3. The left-handed panel shows the relative widths uncorrected for relative width differences, the right-handed panel includes the total width corrections.} 
\label{3gamgam}
\end{figure}

Finally, we look at the implications of the case where the only Higgs boson is the one observed at 125 GeV, that is $h$ and $H$ are nearly degenerate{\footnote{In that case the pseudoscalar $A$ will also have mass 125 GeV, but given the $\beta_0 \simeq 0$ mixing angle in that sector, it will decay invisibly into two neutrinos, not altering the visible branching ratios.}}. We call this boson $h/H$. In that case we have, for the ratio of the number of events in the HTM versus the SM:
\begin{eqnarray}
R_{ XX}=R_{h \to XX}+R_{H \to XX}.
\end{eqnarray}
The values of the masses of  the singly and doubly charged Higgs as functions of $\lambda_5$ remain the same as in Fig. \ref{mhp2} (as they depend only on the $H$ mass).
The plots for the relative signal strength (with respect to the SM one) of $R_{h/H \to \gamma \gamma}$  as a function of $\sin \alpha$, for various values of $\lambda_5$,  are shown in Fig. \ref{3gamgam}, 
in the left panel, for the relative $\gamma \gamma$ widths uncorrected for relative width differences, in the right-handed panels including the total width corrections. At first glance, the results are rather surprising. One would expect that the enhancement from $h \to \gamma \gamma$ will add to the reduction from $H \to \gamma \gamma$ resulting in a perhaps more evenly varying signal, but enhanced with respect to the SM. The fact that this is not the case is apparent from Eq. (\ref{eq:ghHp}). In the degenerate-mass case the term in $\lambda_4$ proportional to $m_h^2-m_H^2$ cancels exactly, and thus from the point of view of the decay into $\gamma \gamma$,  Scenario 3 reproduces exactly the results for  the unmixed case (with $\sin \alpha=0$), see Eq. (\ref{eq:lam4lam2}),  and is approximately independent of $\alpha$. Indeed for most of the parameter space, terms in $m_h^2 \cos^2 \alpha +m_H^2 \sin^2 \alpha \to m_{h/H}^2$, and the same for $\cos \alpha  \leftrightarrow \sin \alpha$.  We checked that the  reduction in the $\gamma \gamma$ signal holds for masses approximately degenerate (within 3-5 GeV){\footnote{This may be relevant as ATLAS and CMS do not agree completely on the  mass of the discovered boson.}}, and gradually becomes an enhancement  for mass splittings of more than $8-10$ GeV.

\subsection{Three-level decays for Scenarios 1, 2 and 3}

\begin{figure}[t]
\center
\begin{center}
$
	\begin{array}{ccc}
\hspace*{-0.7cm}
\includegraphics[width=3.0in,height=3.0in]{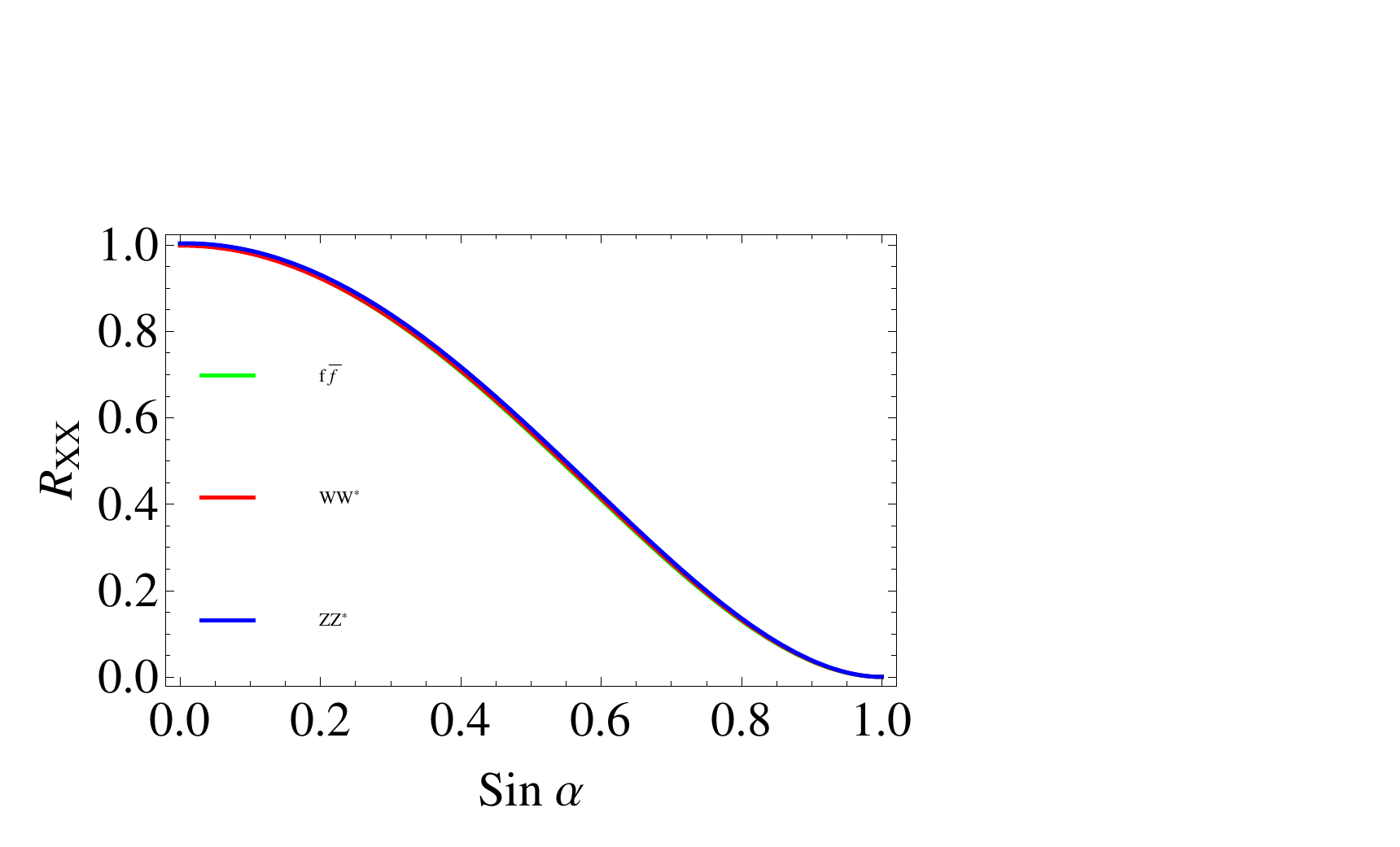}
&\hspace*{-2.0cm}
	\includegraphics[width=3.0in,height=3.0in]{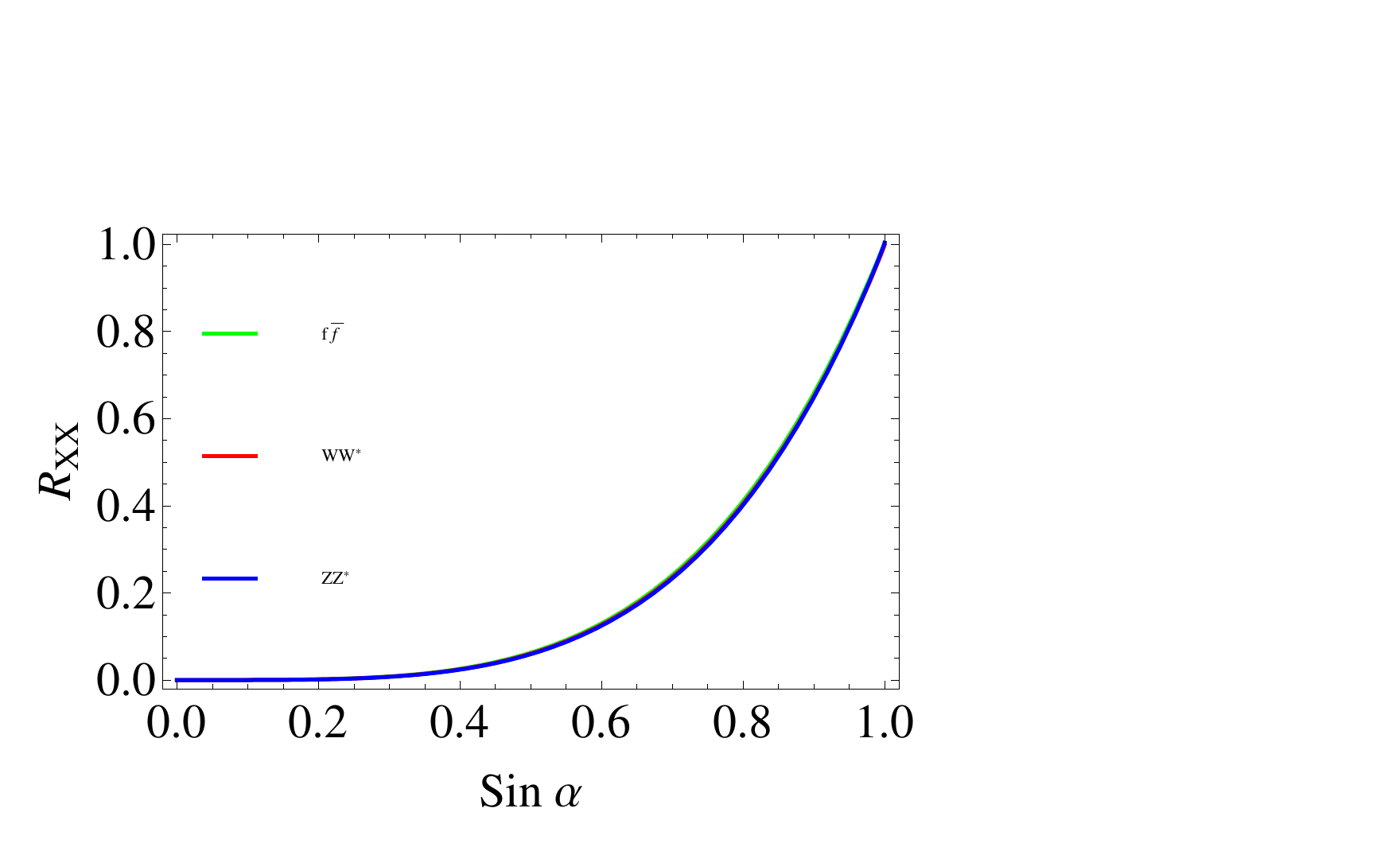}
&\hspace*{-2.0cm}
	\includegraphics[width=3.0in,height=3.0in]{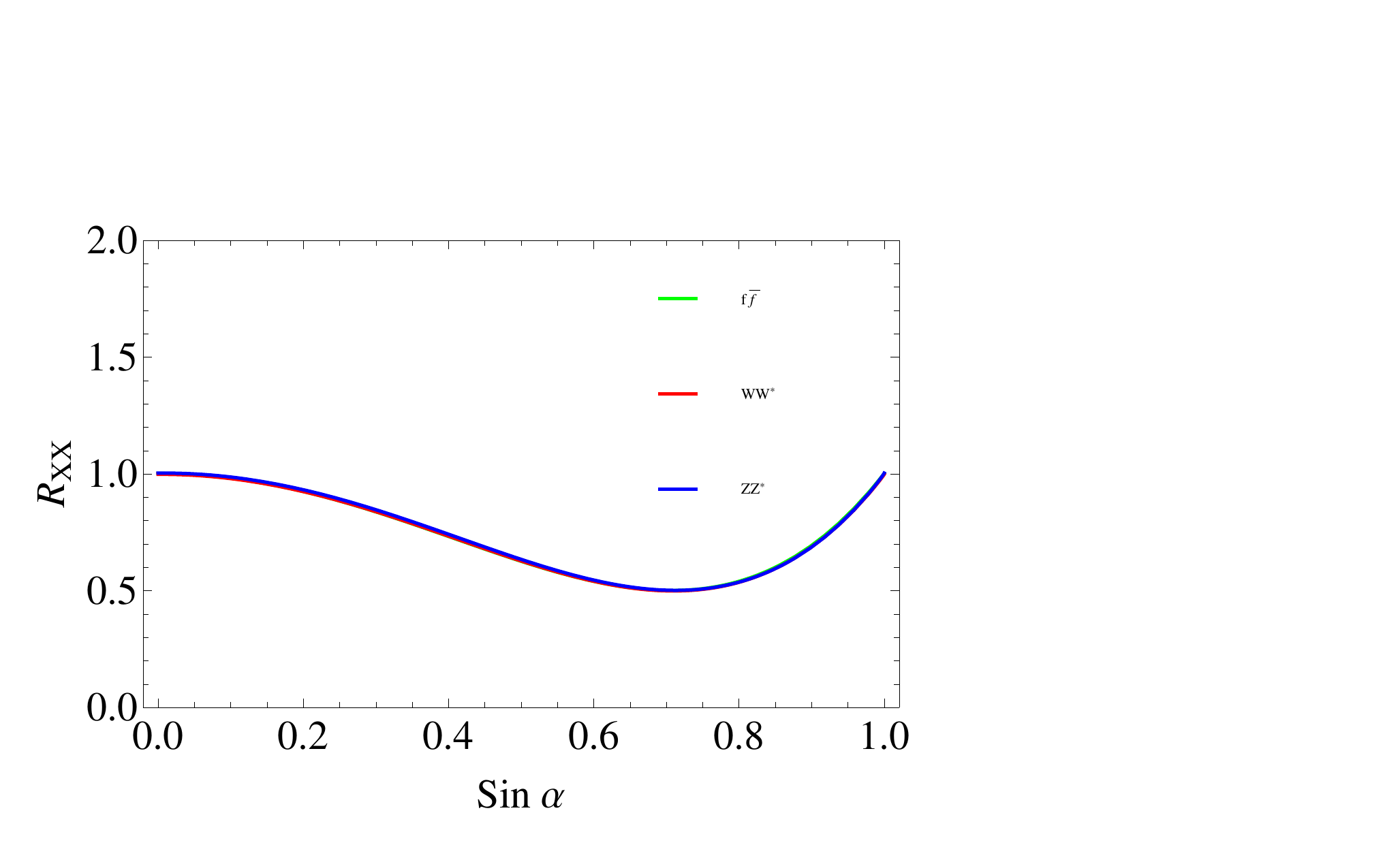}\\
	\hspace*{-0.7cm}
	\includegraphics[width=3.0in, height=3.0in]{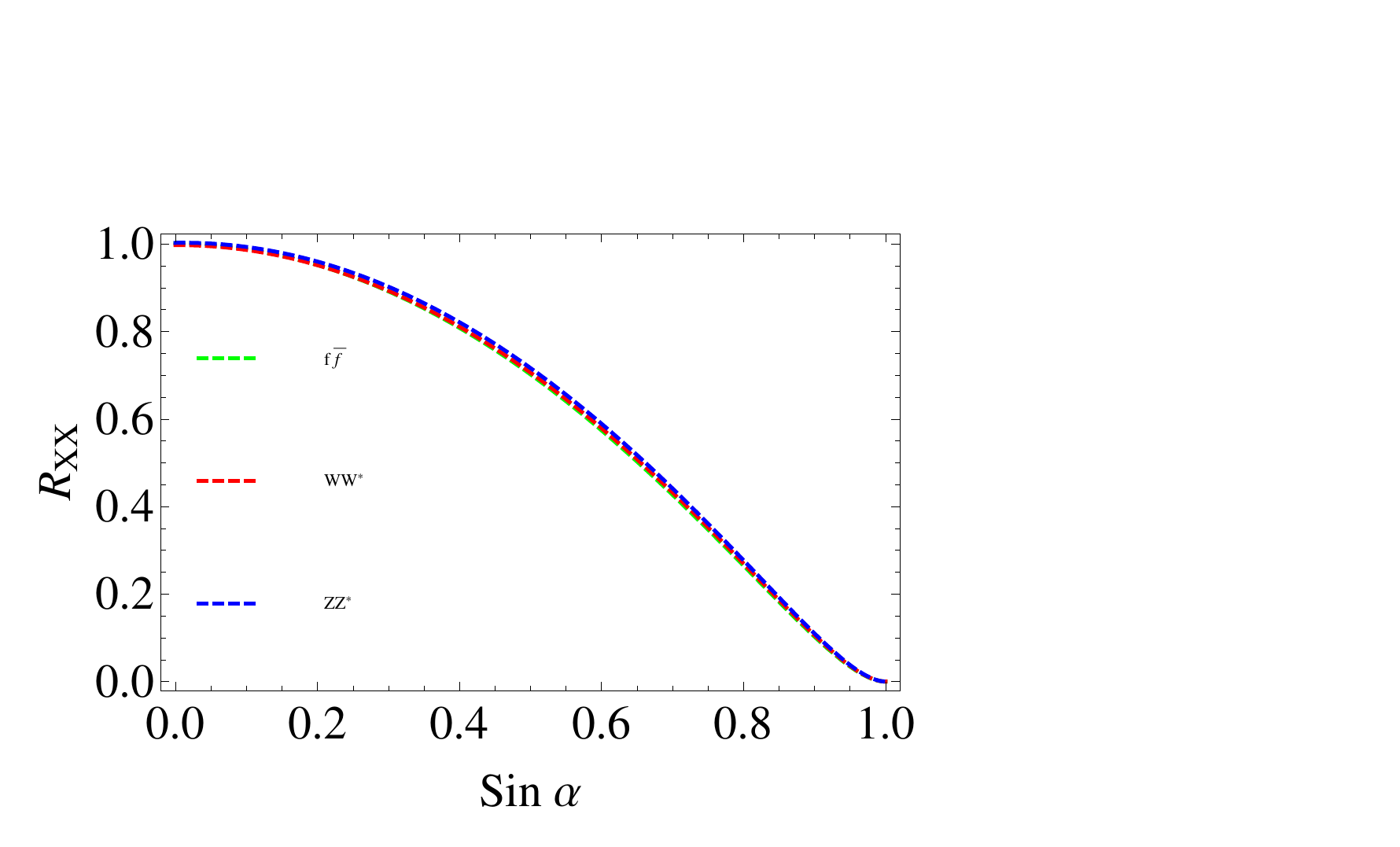}
&\hspace*{-2.0cm}
	\includegraphics[width=3.0in,height=3.0in]{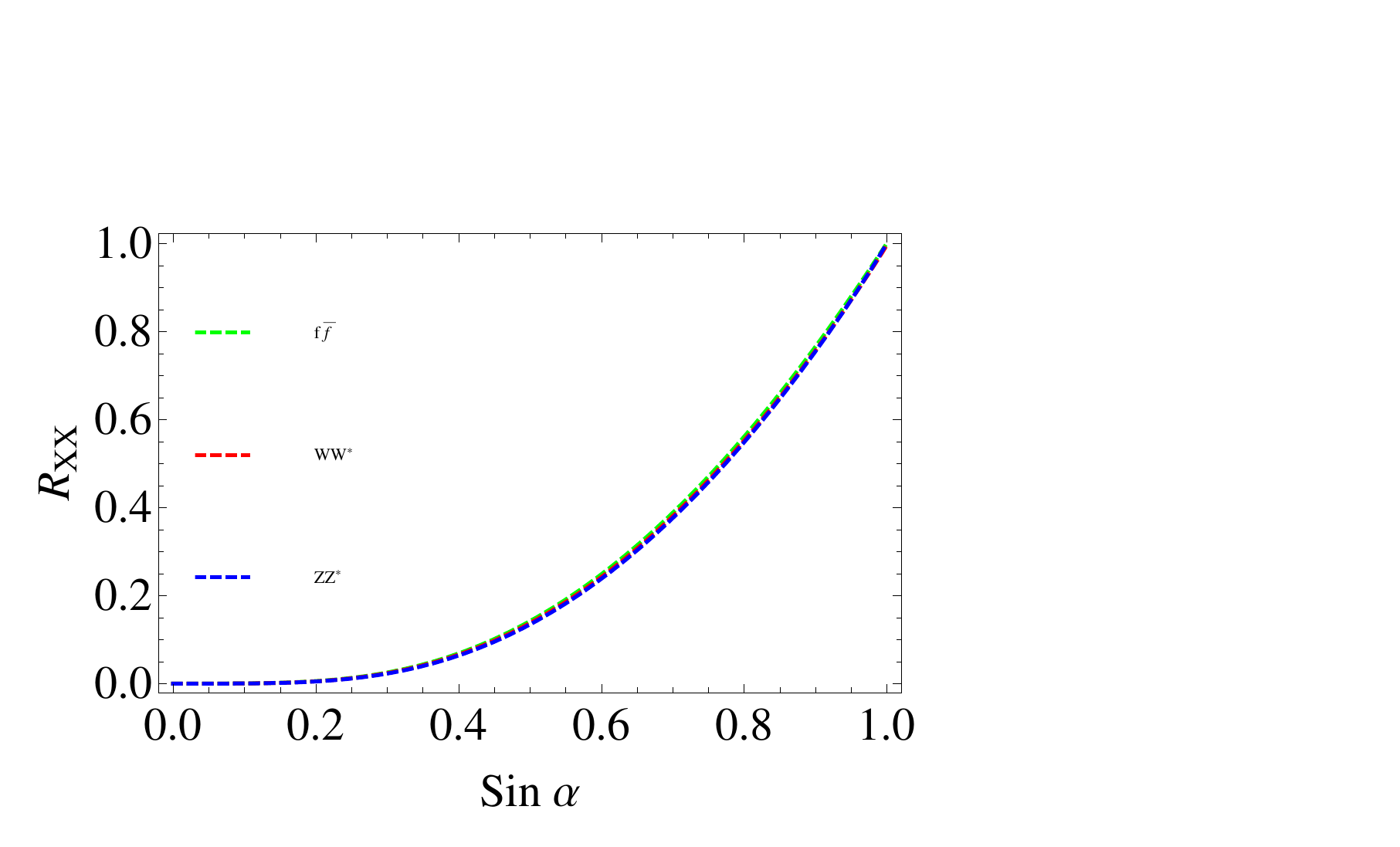}
&\hspace*{-2.0cm}
	\includegraphics[width=3.0in,height=3.0in]{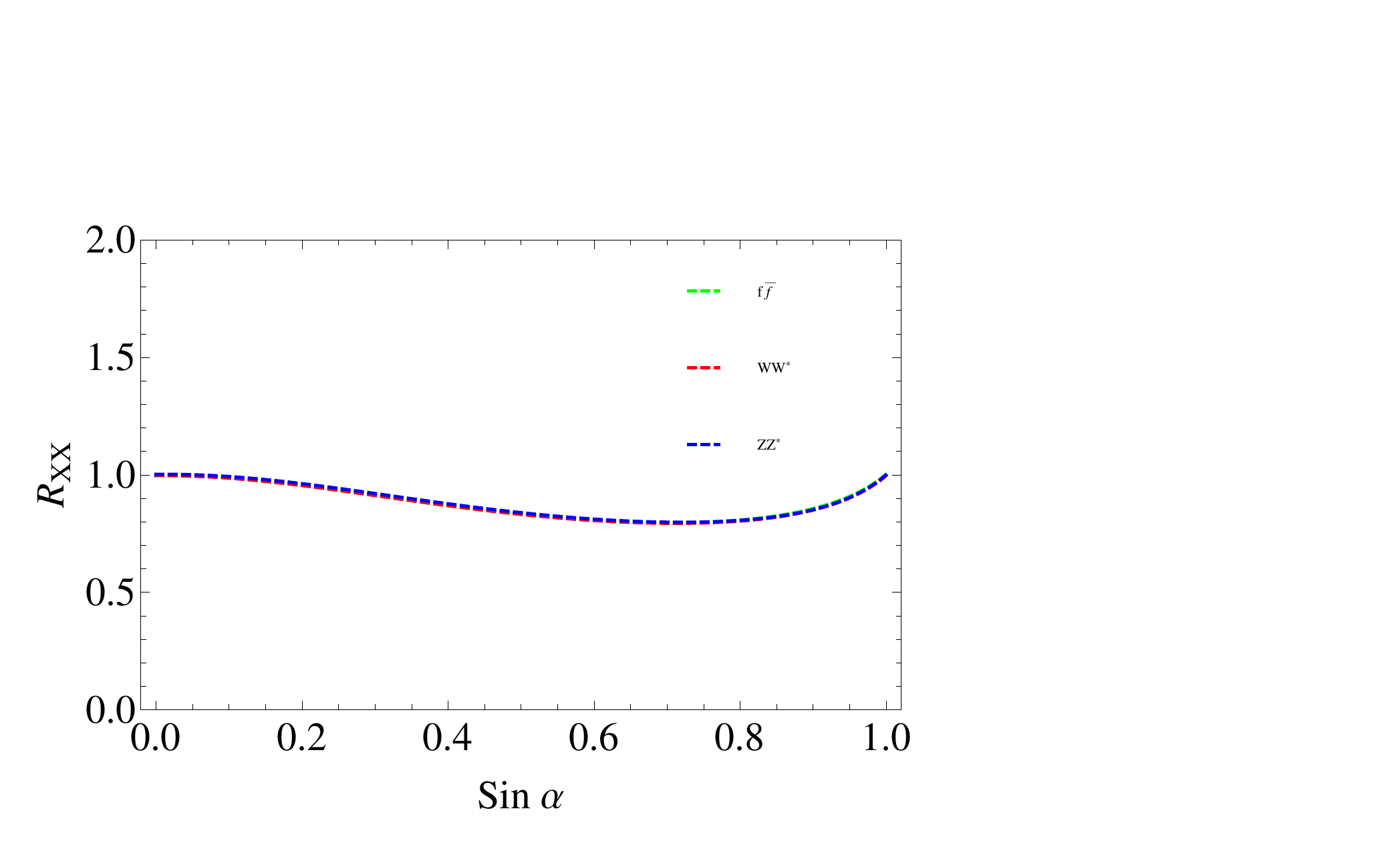}
        \end{array}$
        \end{center}
\caption{Relative Branching Ratios for $h\to f {\bar f}, ~WW^*,~ZZ^*$ in Scenarios 1 and 2  (left column),  for $H\to f {\bar f}, ~WW^*,~ZZ^*$ for Scenarios 1 and 2 (middle column), and for $h,H \to f {\bar f}, ~WW^*,~ZZ^*$ in Scenario 3 (right column)  as a function of the mixing angle in the neutral sector. The top row shows the relative ratios without width corrections, the lower row, including the width corrections. } 
\label{hHtree12}
\end{figure}

We conclude this section with an analysis of the tree level decays ($f {\bar f}~, WW^*,~ZZ^*$) of the neutral Higgs bosons in the 3 scenarios. More precise measurements of these decays, combined with the $\gamma \gamma$ would  constrain the model, as all decays rates depend on very few parameters.
In Fig. \ref{hHtree12} we plot the tree level decays, for all Scenarios, with and without width correction. Note that without correcting for the width, the relative branching ratios are mass independent (thus the same for Scenarios 1 and 2) but they depend on whether the boson is $h$ or $H$. All the tree level branching ratios are suppressed with respect to the same ones in the SM and independent of $\lambda_5$, while  the width-corrected relative decay width are very similar for Scenarios 1 and 2, and thus we show only one. For values of the angles $\alpha$ for which the relative branching ratio to $\gamma \gamma$ falls within the allowed range, the tree-level branching ratios for Scenarios 1 and 2 can lie anywhere between 0.05 and 0.9. Thus more precise measurements of these ratios would give  an indication of the value of the mixing ($\sin \alpha$), which will pick up a definite value of the mass splittings, allowing for a prediction of $m_{H^{++}}$ and  $m_{H^{+}}$. In particular, for Scenario 2, which is favored by the measurements of $h \to \gamma \gamma$ branching ratios, the decays of $H \to f {\bar f}$ obey $0.1< R(H \to b{\bar b})<0.25$ in the region $0.5 < \sin \alpha <0.7$, thus overlapping with regions allowed by the $\gamma \gamma $ constraints for $\lambda_5=-3/2$ and $-2$.  The tree level graphs for Scenario 3, both with and without width correction show that these branching ratios are very close to the SM ones, and relatively independent of $\sin \alpha$, reproducing as before the case for unmixed neutral Higgs bosons. The high branching ratio into $b \bar b$ and $\tau^- \tau^+$  is achieved {\it only} accompanied by a significant reduction in the $\gamma \gamma$ branching ratio. At present, this scenario is disfavored by the measurements at the LHC of $\gamma \gamma$ widths.

\subsection{ Predictions for $H$ and $h$ decay width to $Z\gamma$ }
\label{sec:Zgamma}

As a further test of the implications of the HTM with non-trivial mixing, we evaluate the loop mediated Higgs decay  $h, H \rightarrow Z \gamma$. In the SM the decay $\Phi \to Z \gamma$ is similar to the one for $\gamma \gamma$, but with a smaller rate and a further reduced branching ratio of $Z \to \mu^+ \mu^-$ (or $e^+e^-$).  
 Like the decay to $\gamma \gamma$, it  is sensitive to the presence of charged particles in the loop, and is affected by both their charge and weak isospin. Thus deviations from the SM value could signify beyond the Standard Model physics. The SM contribution for a Higgs state at 125 GeV is very small, $[\Gamma (\Phi \to Z \gamma)]_{SM} \simeq 6 \times 10^{-6}$ GeV, yielding a branching ratio of about $1.5 \times 10^{-3}$  \cite{gainer}, comparable to that of $\Phi \to \gamma \gamma$.  The SM contributions from the $W$ boson and top quark, and the HTM from the additional charged scalars to the decay rate of $h, H$ are given by \cite{Djouadi:2005gi,Carena:2012xa}
\begin{eqnarray}
    \Gamma(\Phi \to Z\gamma)_{SM}&= &\frac{G_F^2 M_W^2 m_h^3 \alpha}{64 \pi^4 } \left(1-\frac{M_Z^2}{m_h^2}\right)^3 \left| {\cal A}_{SM}\right|^2 \ ,\nonumber\\
     \Gamma(h\to Z\gamma)_{HTM}&= &\frac{G_F^2 M_W^2 m_h^3 \alpha}{64 \pi^4 } \left(1-\frac{M_Z^2}{m_h^2}\right)^3 \left|   {\cal A}_{W}(h) + {\cal A}_{t} (h)+ {\cal A}_0^{H^+}(h) + 2 {\cal A}_0^{H^{++}} (h) \right|^2 \ ,\nonumber\\
\end{eqnarray}
where
\begin{eqnarray}
   && {\cal A}_{SM} = \cos \theta_W A_1(\tau^h_W,\sigma_W) + N_c \frac{Q_t (1-4 Q_t \sin^2 \theta_W)}{\cos \theta_W} A_{1/2}(\tau^h_t,\sigma_t) \ ,\nonumber\\
    && {\cal A}_{W}(h) + {\cal A}_{t}(h) =\nonumber \\
     && g_{hWW} \cos \theta_W A_1(\tau^h_W,\sigma_W) +  g_{h t{\bar t}} \, \frac{N_cQ_t (1-4 Q_t \sin^2 \theta_W)}{\cos \theta_W} A_{1/2}(\tau^h_t,\sigma_t) \ ,\nonumber\\
   && {\cal A}^{H^+}_0(h) = \frac {1}{ \sin \theta_W }  g_{ZH^+H^-} {\tilde g}_{hH^+H^-} \ A_0(\tau^h_{H^+},\sigma_{H^+}
    )\ , \nonumber\\
   && {\cal A}^{H^{++}}_0(h) =\frac {1}{ \sin \theta_W}   g_{ZH^{++}H^{--}} {\tilde g}_{hH^{++}H^{--}} \ A_0(\tau^h_{H^{++}},\sigma_{H^{++}}
    )\ ,
\end{eqnarray}
where $\tau^h_i=4m_i^2/m_h^2, ~ \sigma_i=4m_i^2/M_Z^2$,   
$
g_{ZH^{++}H^{--}}=2 \cot 2 \theta_W, ~ g_{ZH^{+}H^{-}}=-\tan \theta_W$, $g_{hWW}$ is given by Eq. (\ref{littleh1WW}),
and ${\tilde g}_{hH^{++}H^{--}}$ and  ${\tilde g}_{hH^{+}H^{-}}$ are given by  Eqs. (\ref{eq:redgcallittlehHp}). The loop functions are given by
\begin{eqnarray}
A_1(\tau,\sigma)&=& 4 (3-\tan^2\theta_W) I_2(\tau,\sigma)+ \left[ (1+2\tau^{-1}) \tan^2\theta_W - (5+2\tau^{-1})\right] I_1(\tau,\sigma) \ , \nonumber\\
 A_{1/2}(\tau, \sigma)& =& I_1(\tau, \sigma)-I_2(\tau, \sigma) \ , \nonumber\\
A_{0}(\tau, \sigma)& =& I_1(\tau, \sigma) \ ,
\end{eqnarray}
where
\begin{eqnarray}
 I_1(\tau, \sigma) &=& \frac{\tau \sigma}{2(\tau-\sigma)} + \frac{\tau^2 \sigma^2}{2(\tau-\sigma)^2}[ f(\tau^{-1})-f(\sigma^{-1})] + \frac{\tau^2 \sigma}{(\tau-\sigma)^2}[g(\tau^{-1})-g(\sigma^{-1})] \ ,\nonumber\\
 I_2(\tau, \sigma) &=& - \frac{\tau \sigma}{2(\tau-\sigma)} [ f(\tau^{-1})-f(\sigma^{-1})]  \ , 
 \end{eqnarray}
where $f(\tau)$ is given in Eq. (\ref{eq:ftau}), and 
\begin{equation}
     g(\tau^{-1}) = \sqrt{\tau-1}\, \arcsin \sqrt{\tau^{-1}} \  ~{\rm for}~\tau>1.
\end{equation}
The decay of $H\to \gamma \gamma$ can be evaluated as before, using the same formulas, with the replacements $h \to H$, $m_h \to m_H$:
\begin{eqnarray}
     \Gamma(H\to Z\gamma)_{HTM} &= &\frac{G_F^2 M_W^2 m_H^3 \alpha}{64 \pi^4 } \left(1-\frac{M_Z^2}{m_H^2}\right)^3 \left|   {\cal A}_{W}(H) + {\cal A}_{t} (H)+ {\cal A}_0^{H^+}(H) + 2 {\cal A}_0^{H^{++}} (H) \right|^2 \ ,\nonumber\\
\end{eqnarray}
where
\begin{eqnarray}
    && {\cal A}_{W}(H) + {\cal A}_{t}(H) =\nonumber \\
    &&g_{HWW} \cos \theta_W A_1(\tau^H_W,\sigma_W)+g_{H t{\bar t}} \, N_c\frac{Q_t (1-4 Q_t \sin^2 \theta_W)}{\cos \theta_W} A_{1/2}(\tau^H_t,\sigma_t) \ ,\nonumber\\
  &&  {\cal A}^{H^+}_0(H) =
 \frac {1}{  \sin \theta_W }  g_{ZH^+H^-} {\tilde g}_{HH^+H^-} \ A_0(\tau^H_{H^+},\sigma_{H^+}
    )\ , \nonumber\\
&&    {\cal A}^{H^{++}}_0(H) =\frac {1}{  \sin \theta_W }  g_{ZH^{++}H^{--}} {\tilde g}_{HH^{++}H^{--}} \ A_0(\tau^H_{H^{++}},\sigma_{H^{++}}
    )\ ,
\end{eqnarray}
where $\tau^H_i=4m_i^2/m_H^2, ~ \sigma_i=4m_i^2/M_Z^2$, $g_{HWW}$ is given in Eq. (\ref{BigH1WW}), 
and ${\tilde g}_{HH^{++}H^{--}}$ and  ${\tilde g}_{HH^{+}H^{-}}$ are given by  Eqs. (\ref{eq:redgcalBigHHp}).

\begin{figure}[t]
\center
\vskip -0.4in 
\begin{center}
$
	\begin{array}{cc}
\hspace*{-0.7cm}
\includegraphics[width=3.7in,height=3.6in]{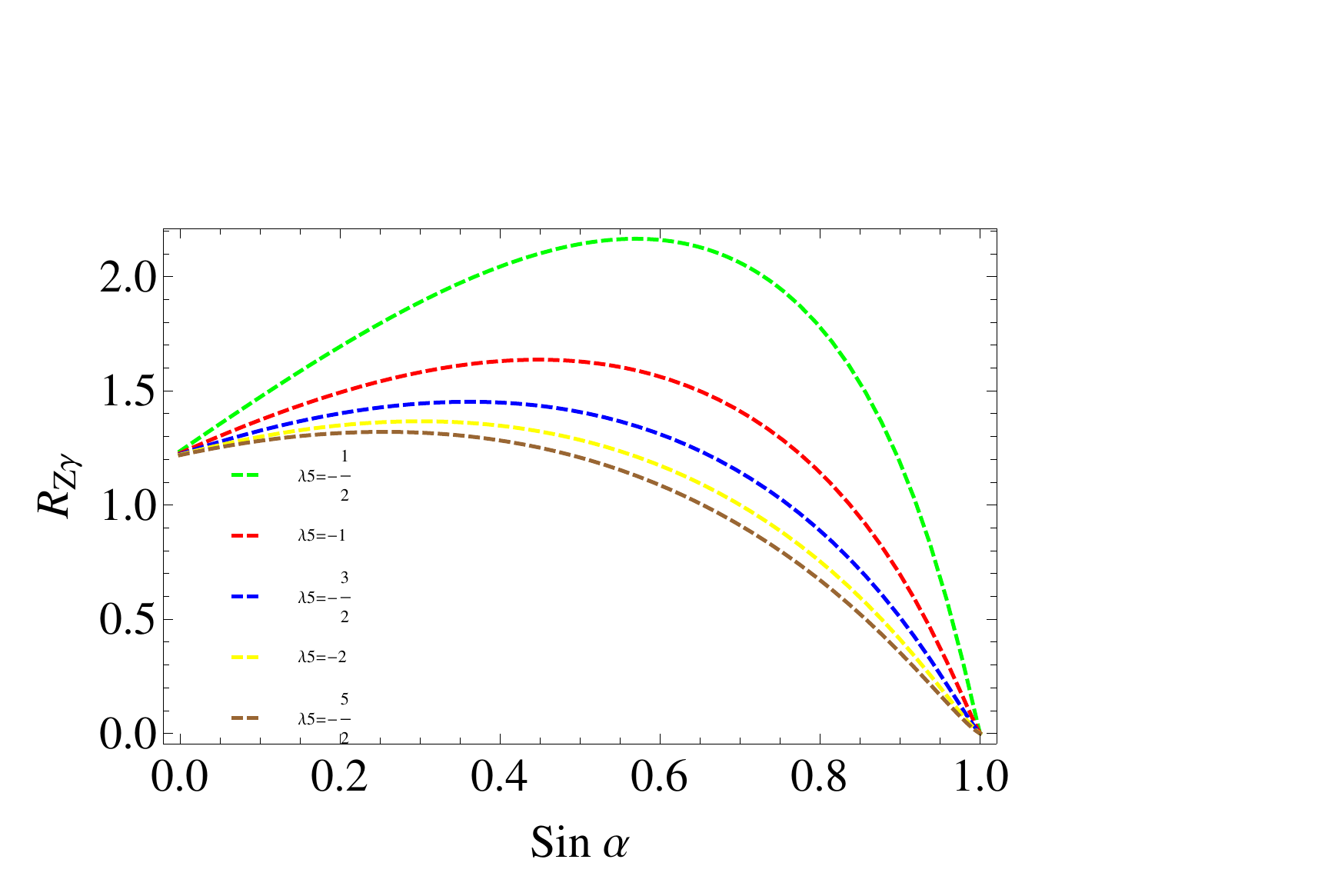}
&\hspace*{-1.4cm}
	\includegraphics[width=3.8in,height=3.7in]{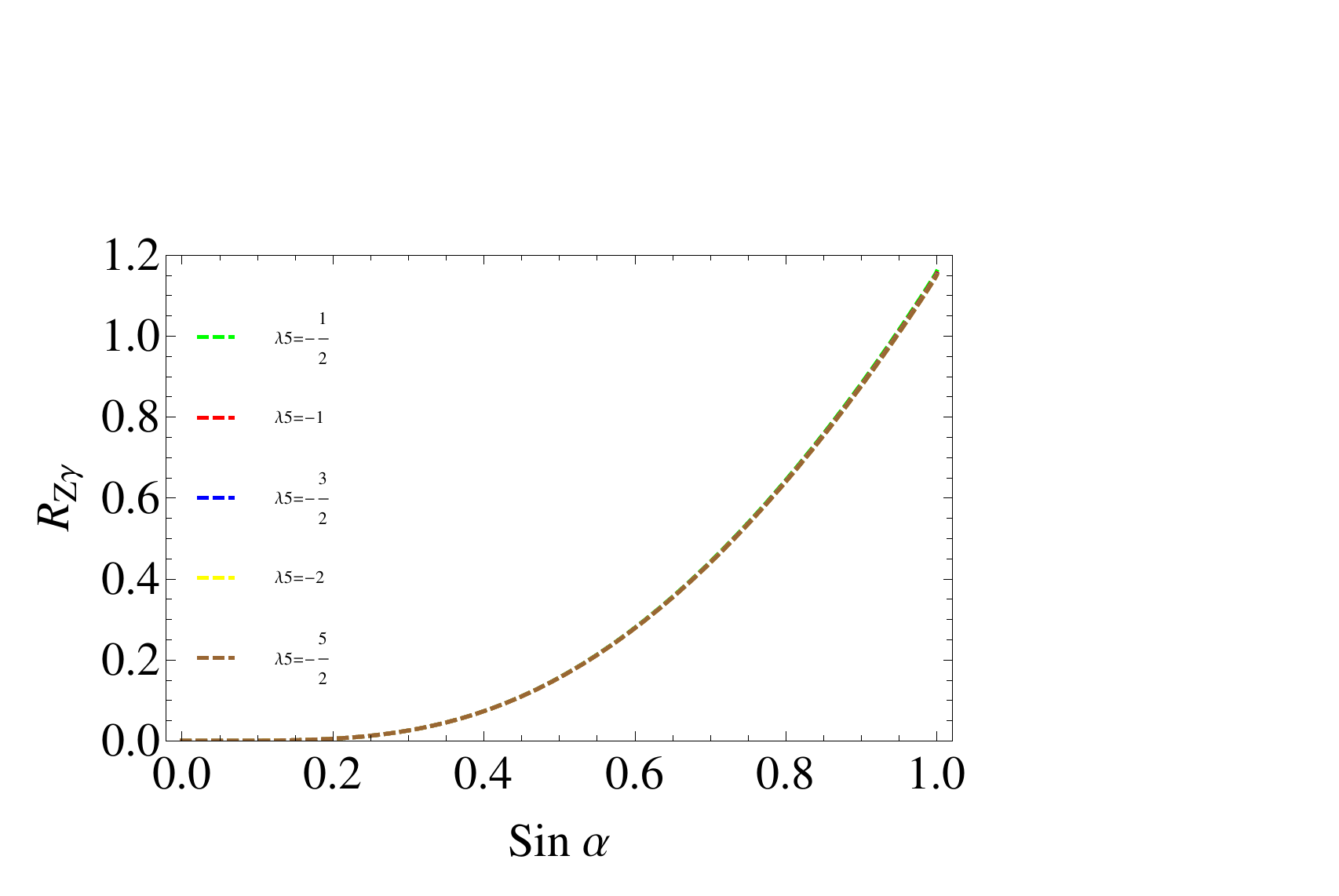}\\
\hspace*{-0.7cm}
	\includegraphics[width=3.7in,height=3.6in]{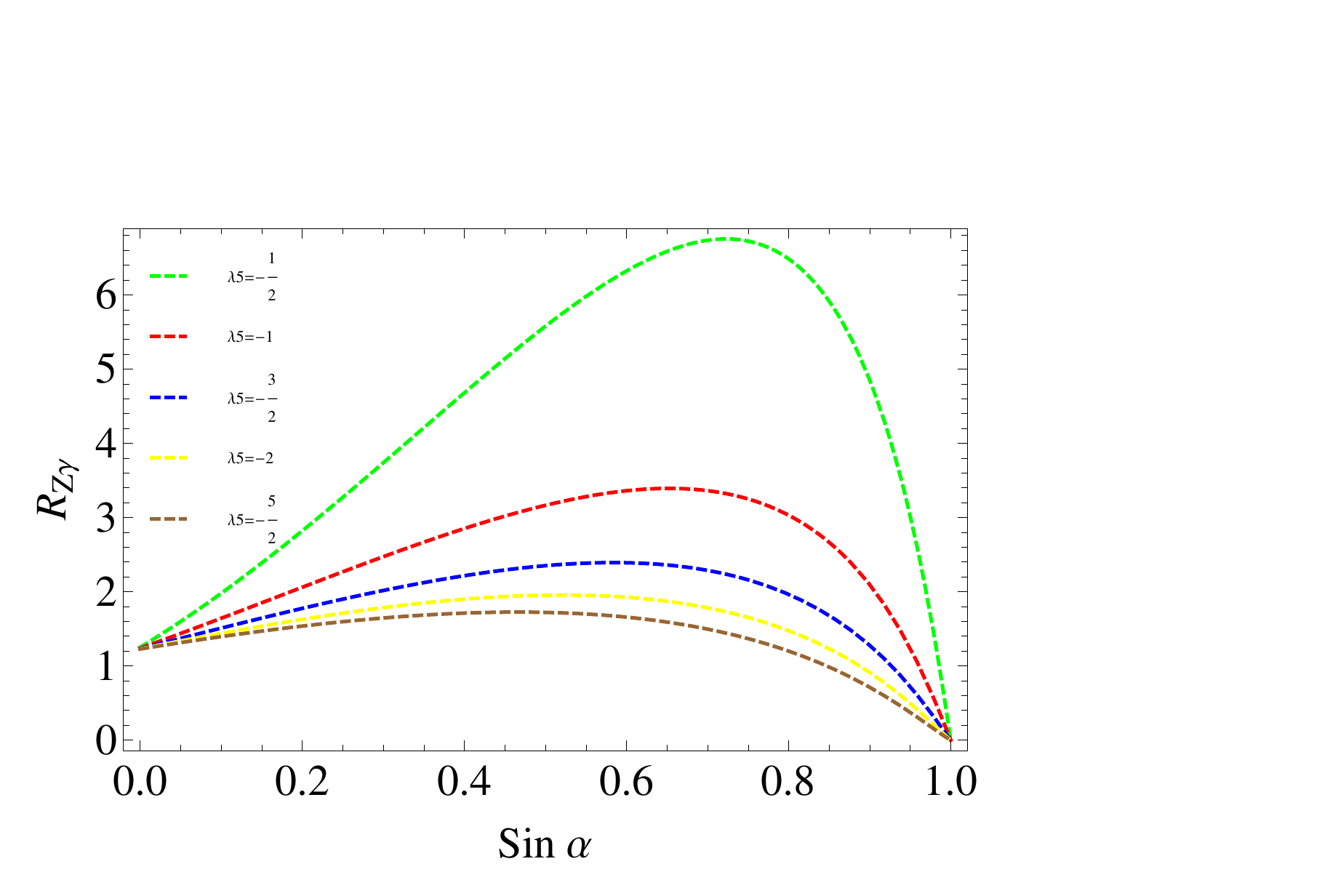}
&	\hspace*{-1.4cm}
	\includegraphics[width=3.8in, height=3.7in]{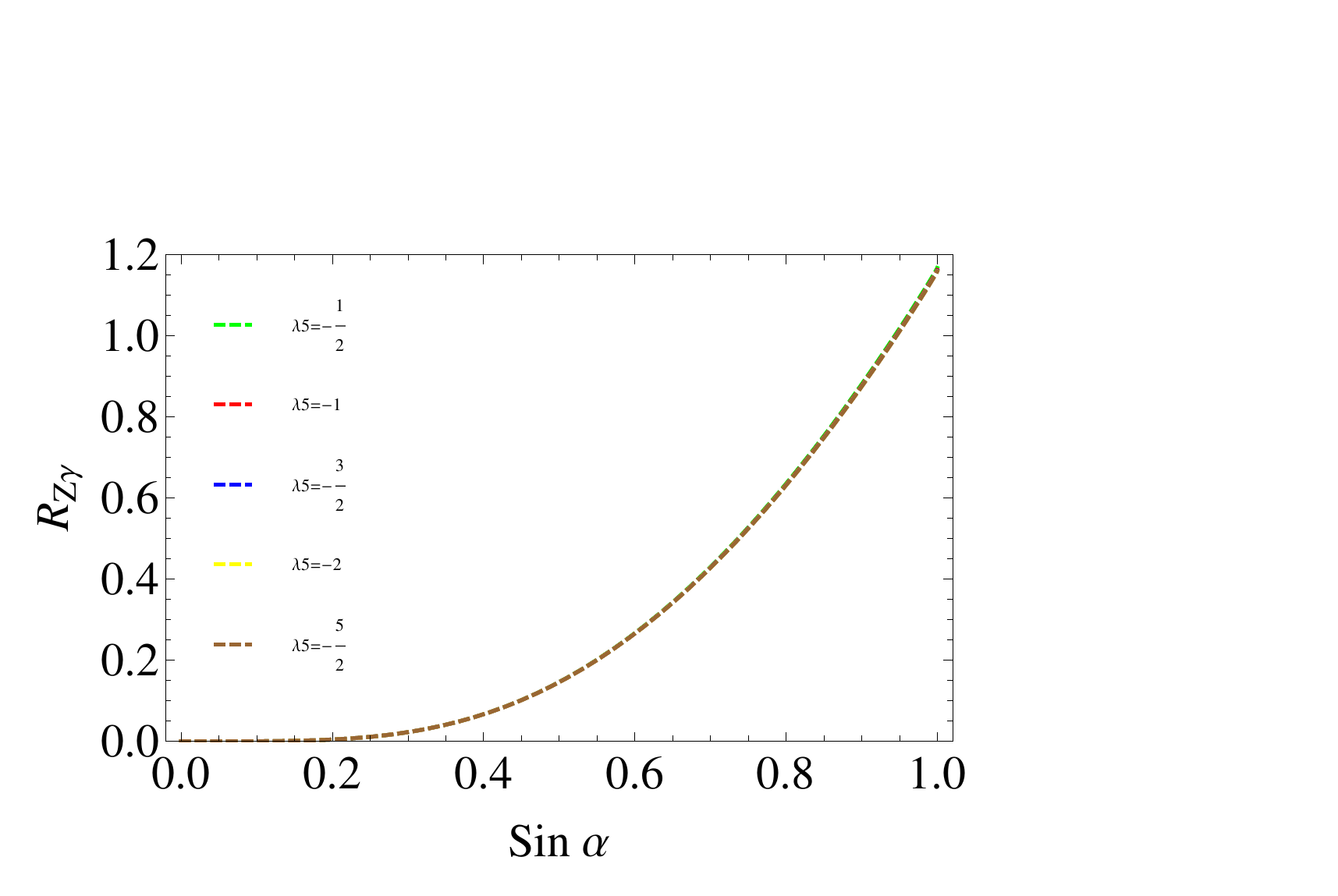}
	\end{array}$
        \end{center}
\caption{Relative Branching Ratios for $h \to Z \gamma$ (left column) and $H \to  Z \gamma$ (right column) with width corrections for Scenario 1(upper row), and Scenario 2 (lower row) as a function of  $\sin \alpha$, for various $\lambda_5$ values.} 
\label{Zgam}
\end{figure}

Comparison with the SM predictions lead to the modification factor $R_{Z\gamma}$ for the $h,H \rightarrow Z \gamma$ decay rate with 
\begin{eqnarray}
R_{h,H \to Z \gamma}&\equiv& \frac {\sigma_{\rm HTM}(gg \to h,H \to Z \gamma )}{\sigma_{\rm SM}(gg \to \Phi \to Z \gamma )} \nonumber \\
&=&\frac{[\sigma( gg \to  h, H) \times \Gamma(h,H \to Z \gamma)]_{HTM}}{[\sigma( gg \to \Phi) \times \Gamma(\Phi \to Z \gamma )]_{SM}} \times \frac{[\Gamma(\Phi)]_{SM}}{[\Gamma(h,H)]_{HTM}}\ .
\end{eqnarray}
In Fig.~\ref{Zgam} we plot the relative width factor $R_{Z\gamma}$ as a function of the scalar mixing $\sin \alpha$ for Scenarios 1 and 2, and for various mass splittings in the charged sector. One can see that the model predicts an enhancement in $h \to Z \gamma$, inherited in part from the enhancement in $\gamma \gamma$ due to ${\tilde g}_{hH^{++}H^{--}}$ and ${\tilde g}_{hH^{+}H^{-}}$. This is the case especially for relatively light charged and doubly charged Higgs masses, and can reach a factor of 3 in Scenario 2, for the region favored by the $\gamma \gamma$ decay measurement. Unlike other signals, even the lighter $H$ can show a modest enhancement in $Z\gamma$, but for values of $\sin \alpha=0.9 \to 1$, and this enhancement is independent of $\lambda_5$ values. A measurement of the rare decay into $Z \gamma $ could thus serve as a confirmation of this scenario in HTM. 

\begin{figure}[t]
\center
\vskip -0.4in 
\begin{center}
\hspace*{1.4cm}
	\includegraphics[width=3.8in,height=3.8in]{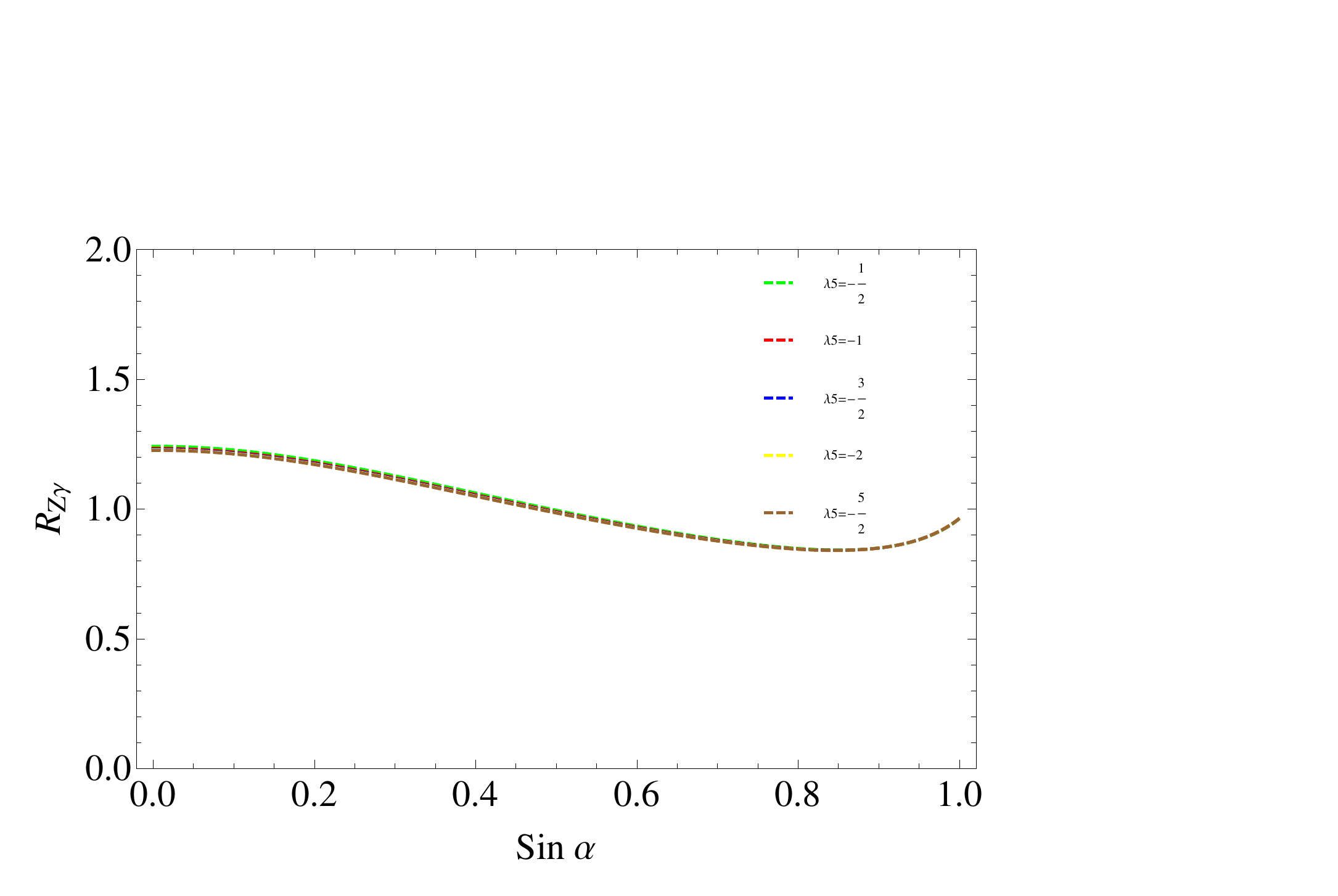}
 \end{center}
\caption{Relative Branching Ratios for $h/H \to Z \gamma$ in Scenario 3 as a function of  $\sin \alpha$, for various $\lambda_5$ values.} 
\label{Zgam3}
\end{figure}

Even for Scenario 3, the HTM predicts some modest enhancement over the SM, and this enhancement is independent of $\lambda_5=m_{H^{++}}^2-m_{H^+}^2$ but is valid for small mixings $\sin \alpha \in (0, ,0.2)$. The results are shown in Fig \ref{Zgam3}, where we only plot the relative branching ratios corrected  for the width. The variations with $\sin \alpha$ are very small, and more pronounced for the uncorrected relative width, overall similar to those for $\gamma \gamma$, and indicative of the effects of the charged Higgs bosons in the loop.

\newpage
\section{Conclusions}
\label{sec:conclusion}
We presented a comprehensive analysis of the decay ratios on the CP-even neutral Higgs bosons in the HTM, allowed to mix with arbitrary angle $\alpha$. Of the bare states in the model, one is the usual neutral component of the SM Higgs doublet, the other is the neutral component of a Higgs triplet, introduced to provide neutrino masses. 
We studied the  ratios of production and decay of the Higgs in this model at tree and one-loop level, relative to the ones in the SM. We have shown that, in the case where the two Higgs do not mix, positivity conditions on the scalar potential forbid an enhancement of the branching ratio into $\gamma \gamma$. Allowing for arbitrary mixing, these conditions require that $h$ (the neutral Higgs which is the corresponding SM one in the no mixing limit) is heavier than $H$. We have also shown that, if the Higgs are allowed to mix non-trivially, the relative branching ratio into $\gamma \gamma$ of $h$ with respect to the SM Higgs  can be enhanced, and that, for all these cases, the singly-charged Higgs boson is lighter than the doubly-charged boson, and both are heavier than $H$. This is a very different scenario from the unmixed one, where $h$ is the lighter neutral Higgs, and the doubly-charged Higgs bosons are lighter than the singly charged Higgs, who in turn are lighter than the neutral triplet $H$.

We allowed the mixing angle $\alpha$ to vary and expressed all the  couplings in the Higgs potential as a function of this angle, and of the square-mass splitting $\lambda_5$. We analyzed three scenarios. The first one, where $H$ is the boson observed at 125 GeV and $h$ is the CMS excess at $136$ GeV, is disfavored by the data, as the branching ratio of $H\to \gamma \gamma$ is always reduced with respect to SM expectations. However, Scenario 2, where $h$ is the boson observed at 125 GeV, and $H$ the excess observed in $e^+e^-$ at LEP at 98 GeV, is favored by the data, and consistent with all other measurements. This scenario can also explain a lighter Higgs $H$ which is missed by colliders because of significantly reduced decay into $\gamma\gamma$.  In both of these scenarios the tree-level decay rates of $h$ and $H$ are reduced with respect to the SM. Should such a reduction survive more precise measurements, Scenario 2  looks very promising.
The case where the two neutral bosons are (almost) degenerate resembles very much the unmixed neutral case. The relative branching ratio into $\gamma \gamma $ is suppressed, and even if the tree-level decays are at the same level as expected in the SM, this scenario is disfavored at present by the LHC measurements.

Finally, we have tested all scenarios with the decay  $h, H \to Z \gamma$ and we find significant enhancements, relevant especially for Scenario 2, which shows enhancements in $\gamma \gamma$ for the boson at 125 GeV; and even for Scenario 3, in which the two Higgs bosons are (almost) degenerate.  As this branching ratio is also sensitive on the extra charged particles in the model, a precise measurement could shield some light on the structure of the model.

In conclusion, the power to discriminate the SM Higgs boson from Higgs bosons in extended models depends critically on differentiating their couplings and decays.   We have shown that a very simple model, in which only one extra (triplet) Higgs representation is added to the SM to allow for neutrino masses, shows promise in being able to explain the present data at LHC, and indicated how, with more precise data, this Higgs sector can be validated or ruled out.

\acknowledgments
The work of S. B. and M.F.  is supported in part by NSERC under grant number SAP105354. 


\end{document}